\documentclass[a4paper,11pt]{article}
\usepackage{jcappub} 
\usepackage{graphicx}
\usepackage{amssymb,amsmath}
\usepackage{subfiles}
\usepackage{hyperref}
\usepackage{footmisc}
\usepackage{booktabs}
\usepackage{lineno}
\usepackage[utf8]{inputenc}
\usepackage{graphicx}
\usepackage{xcolor,bm}
\usepackage[T1]{fontenc}
\usepackage[english]{babel}
\usepackage[utf8]{inputenc}

\usepackage{lineno}
\usepackage{soul}
\usepackage{bm}
\usepackage{subfig}
\usepackage{hyphenat}
\usepackage{seqsplit}

\title{\boldmath Sensitivity 
to  sub-GeV dark matter from cosmic-ray scattering with very-high-energy gamma-ray observatories}

\author[a,b]{Igor Reis\,\,}
\emailAdd{igorreis@ifsc.usp.br}\emailAdd{igor.reis@cea.fr}

\author[b]{, Emmanuel Moulin\,\,}
\emailAdd{emmanuel.moulin@cea.fr}

\author[a]{, Aion Viana\,\,}
\emailAdd{aion.viana@ifsc.usp.br}

\author[c,d]{,\\ Victor P. Goncalves}
\emailAdd{barros@ufpel.edu.br}

\affiliation[a]{Universidade de São Paulo, Instituto de Física de São Carlos, Av. Trabalhador São Carlense 400, São Carlos, Brazil}
\affiliation[b]{Irfu, CEA Saclay, Université Paris-Saclay, F-91191 Gif-sur-Yvette, France}
\affiliation[c]{Institute of Physics and Mathematics, Federal University of Pelotas, \\
  Postal Code 354,  96010-900, Pelotas, RS, Brazil}
\affiliation[d]{Institute of Modern Physics, Chinese Academy of Sciences, Lanzhou 730000, China}

\abstract{
Huge efforts have been deployed to detect dark matter (DM) in the GeV-TeV mass range involving various detection techniques, and led to strong constraints in the available parameter space. We compute here the sensitivity to sub-GeV DM that can be probed from the inevitable cosmic-ray scattering onto DM particles populating the Milky Way halo. Inelastic scattering of energetic cosmic rays off DM would produce high-energy gamma rays in the final state, providing a new avenue to probe the poorly-constrained so far sub-GeV dark matter mass range. In this work we derive 
sensitivity forecasts for the inelastic cosmic-ray proton - DM cross section  for current and future very-high-energy gamma-ray observatories such as H.E.S.S., LHAASO, CTA and SWGO in the 100 eV to 100 MeV mass range. These inelastic cross section constraints are converted to the elastic proton - DM cross section  
to highlight further complementarity with cosmological, collider and direct detection searches. 
The sensitivity computed at 95\% confidence level on the elastic cross section reaches $\sim$2$\times$ 10$^{-32}$ cm$^2$ for a 100 keV DM mass for H.E.S.S.-like and $\sim$7$\times$ 10$^{-34}$ cm$^2$ for a $\sim$1 keV DM mass for LHAASO. The sensitivity prospects for CTA and a strawman SWGO model reach $\sim$6$\times$ 10$^{-34}$ cm$^2$ and $\sim$4$\times$ 10$^{-35}$ cm$^2$, for DM masses of 10 keV and 1 keV, respectively. The sensitivity reach of the gamma-ray observatories considered here enables to probe an uncharted region of the DM mass - cross section parameter space. 
}

\begin{document}
\maketitle
\flushbottom

\section{Introduction}
\label{sec:intro}
Dark Matter (DM) pervades the universe at all scales revealing its presence through gravitational interaction. A substantial body of evidence, such as galaxy rotation curves, galaxy cluster collisions, the Cosmic Microwave Background (CMB) and large-scale structure formation, implies that most of the matter in the universe does not behave as the known baryonic matter~\cite{Kolb:1990vq,Clowe:2006eq,Garrett:2010hd,Dodelson:2011qv}.
Various approaches have been deployed to probe the non-gravitational interaction of DM such as its production at colliders~\cite{Kahlhoefer:2017dnp}, its scattering 
off target nucleons or electrons in underground detectors~\cite{Schumann:2019eaa}, commonly referred to as direct DM detection, or its decays or annihilations probed by indirect detection~\cite{Strigari:2018utn}. Strong constraints have been derived in the GeV-TeV mass regime from direct and indirect detection experiments. Indirect searches in gamma rays set strong constraints on the velocity-weighted annihilation cross-section 
probing the natural scale expected for thermal-relic Weakly interacting massive particles (WIMPs)~\cite{Ackermann:2015zua,HESS:2022ygk}.
 For DM particle $\chi$ with masses $m_\chi$ above a few ten GeV, current constraints probe DM-nucleon cross-section expected for electroweak DM models~\cite{XENON:2023cxc}. However, the constraints are dramatically relaxed, by orders of magnitude, for DM particle with sub-GeV masses. For DM particle velocity of $\sim$10$^{-3} c$ as expected in the Milky Way (MW) halo, sub-GeV DM would produce nucleon recoils at keV energies where standard direct detection experiments lose sensitivity. Measurements of electronic recoils would be able to probe DM masses down to about 1 MeV, see, for instance, Ref.~\cite{XENON:2021qze}. 
 To explore these sub-GeV masses, current experimental endeavors have shifted focus towards alternative targets and instrumentation techniques~\cite{Battaglieri:2017aum,Ambrosone:2022mvk,Elor:2021swj}. Lately, an alternative method for detecting light DM has been explored in the context of direct detection experiments, which involves observing DM particles that have obtained relativistic energies via the elastic collision with cosmic rays (CRs), constituting what has been called cosmic-ray \textit{boosted dark matter}, see, for instance, Ref.~\cite{Bringmann:2018cvk,Agashe:2014yua}. This implies standard direct detection experiments to be able to detect light DM particles, due to the induced much larger recoil energies. 
Assuming fermionic DM, kinematic measurements in dwarf spheroidal galaxies 
set model-independent lower bounds on DM masses down to  $\sim$0.1 keV depending on the production scenario~\cite{Alvey:2020xsk, Boyarsky:2008ju}. 
In what follows, no specific DM production scenario is assumed. Should one consider thermally-produced sub-GeV DM, CMB and BBN constraints would be unavoidable (see, for instance, Refs.~\cite{Boehm:2013jpa,Sabti:2019mhn,Depta:2019lbe}).

\begin{figure}[t]
    \centering
    \includegraphics[scale = 0.8]{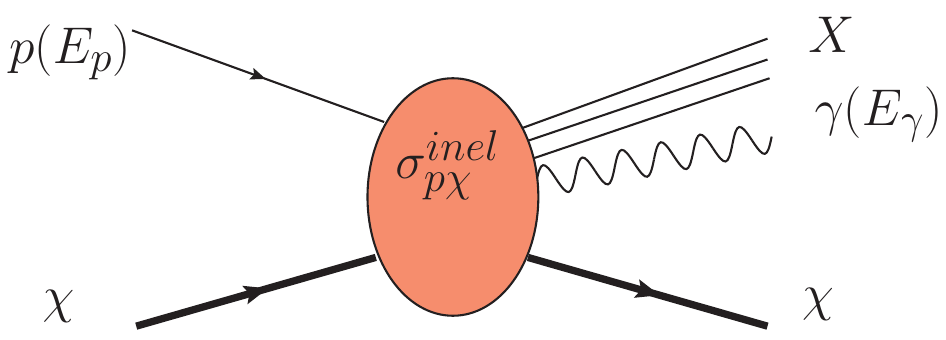}
    \caption{Production of a gamma ray with energy $E_{\gamma}$ in inelastic CR proton - dark matter ($p\chi$) interactions at a center - of - mass energy $\sqrt{s}$. In the DM rest frame, one has that $\sqrt{s} = (2 E_p m_{\chi} + m_p^2 + m_{\chi}^2)^{\frac{1}{2}}$, where $E_p$ is the incident CR proton energy and $m_{\rm p}$ and $m_{\chi}$ are the proton and DM particle mass, respectively. The interaction is described by the inelastic $p\chi$ cross-section, $\sigma^{\rm inel}_{\rm p\chi} (\sqrt{s})$, which is energy dependent in the general case.}
    \label{fig:diagram}
\end{figure}

Inelastic collisions between CR protons and DM particles in the MW will inevitably produce gamma rays. The basic idea, represented in Fig.~\ref{fig:diagram},  is that for CRs of energy $E_{\rm p}$ colliding with DM at large center-of-mass energies $\sqrt{s}$, the protons can be excited or break-up, producing gamma rays of energy $E_{\gamma}$ in the final state. Over the last decades, several models have been proposed to describe the proton - DM interactions, 
considering different assumptions for the couplings and properties of the mediator (see, \textit{e.g.}, Refs.~\cite{Soper:2014ska,Guo:2020oum,Chu:2020ysb,Alvey:2022pad}). Alternatively, one can derive model-independent upper bounds on the inelastic cross-section $\sigma^{\rm  inel}_{\rm p \chi} (\sqrt{s})$, which should be satisfied by all models that intend to describe the inelastic proton - DM interaction.  As the energy dependence of the inelastic $p \chi$ cross-section is still an open question, we will  investigate how such a dependence impacts our findings assuming that $\sigma^{\rm inel}_{\rm p \chi} (\sqrt{s})$ depends on the energy similarly to that  observed in 
proton-proton, meson-proton and real photon-proton collisions at high energies~\cite{Donnachie:2002en}. 
Energetic cosmic-ray protons in the MW would consequently produce diffuse very-high-energy (VHE, E$\gtrsim$100 GeV) gamma rays that could be eventually detected by current and future VHE gamma-ray observatories,  such as H.E.S.S.~\cite{HESS}, LHAASO~\cite{LHAASO}, MAGIC~\cite{MAGIC}, 
VERITAS~\cite{VERITAS}, CTA~\cite{CTA, CTAConsortium:2017dvg} and SWGO~\cite{SWGO}, and further used to place limits  on the inelastic cross-section. In this context, observations of the Galactic Center (GC) and the Inner Galaxy are of prime interest to search for such signatures, as both the CR flux and DM density are measured to be higher there than anywhere in the MW.

In recent years, the elastic proton-DM interaction strength has been constrained severely by various direct detection experiments, as well as by cosmological bounds. In this paper, inspired by Ref.~\cite{Cyburt:2002uw}, we will investigate if the bounds derived for $\sigma^{\rm inel}_{\rm p \chi}$ can impose additional limits on the elastic $p\chi$ cross-section ($\sigma^{\rm el}_{\rm p \chi}$). We will explore such a possibility considering distinct approaches to relate these quantities and the constraints will be recast on the elastic cross-section on nucleon to highlight further complementary with cosmological, collider and direct detection searches.

The paper is organized as follows. Section~\ref{sec:crdm} describes the expected gamma-ray signals from inelastic scattering of high-energy cosmic rays off dark matter particles populating the MW halo, together with gamma-ray backgrounds. 
In section~\ref{sec:observatories} we present current and future VHE gamma-ray observatories and their capabilities used to compute sub-GeV DM sensitivity. 
Section~\ref{sec:stats} outlines the statistical analysis
framework used to compute the sensitivity to proton-DM cross-section. Results and discussion are presented in Section~\ref{sec:results} in light of current constraints on the elastic cross-section between dark matter and proton in the sub-GeV DM mass range. Finally, in Section~\ref{sec:sum} we summarize our main results and conclusions.

\section{Gamma-ray production from cosmic-ray scattering off dark matter}
\label{sec:crdm}

\subsection{Expected VHE gamma-ray signal}
\label{subsec:model}
Gamma rays of energy $E_{\gamma}$ can be produced in inelastic collisions between CR protons and DM particles of mass $m_{\rm \chi}$ populating the MW DM halo.
The associated gamma-ray flux is a function of the 
cross-section of the inelastic interaction between CR protons and DM, $\sigma^{\rm inel}_{\rm p \chi}$, and can be calculated given the functional form of the cosmic ray flux,  $d\Phi_{\rm p}/dE_{\rm p}$, and the DM distribution $\rho_{\rm DM}(r)$ in the MW. 
In particular, in the $\delta$-function approximation~\cite{1996A&A...309..917A} and for a power-law proton flux,  the expected energy-differential flux in gamma rays in a solid and angle $\Delta\Omega$  writes as 
\begin{equation}
\begin{aligned}
    \frac{d\Phi_\gamma (E_{\gamma})}{dE_{\gamma}} = \frac{4 (f_\pi)^{\Gamma-1}}{\Gamma^2}\times \frac{\sigma_{\rm p \chi}^{\rm inel}(E_\gamma/f_\pi, m_\chi)}{m_{\chi}}\times \Phi_0 (E_\gamma/{\rm 10 TeV})^{-\Gamma}
    \times D(\Delta\Omega) \, ,
    \label{eq:flux}
\end{aligned}
\end{equation}
where $\Gamma$ and $\Phi_0$ are the spectral index and normalization of the power-law CR proton flux, respectively, and the factor 
$f_{\pi}$ characterizes the fraction of the proton energy transferred to the secondary neutral pions that decay into gamma rays. As in Ref.~\cite{1996A&A...309..917A}, we will assume in the following $f_{\pi} = 0.17$.  
The gamma-ray production for the interaction between a CR proton and a DM particle of mass $m_{\chi}$ requires a minimum CR proton energy to produce a pion  given by $E^{\rm min}_{\rm p} = (m_{\rm p} + m_{\rm \pi^{0}})\left[1 + m_{\rm \pi^{0}}(2m_{\rm p} + m_{\rm \pi^{0}})/(2m_{\rm \chi}(m_{\rm p} + m_{\rm \pi^{0}}))\right]$, that, 
therefore, implies a minimum value for the gamma-ray energy of $E_{\gamma}^{\rm min} \simeq E_{\rm p}^{\rm min}/f_{\rm \pi}$ at VHE~\cite{Stecker:1971ivh,Gaisser:2016uoy}.
Finally,  the $D$-factor can be defined as the integral of DM density in the MW halo along  the line-of-sight  (l.o.s.) $s$ and over the solid angle $\Delta\Omega$ as
\begin{eqnarray}
D(\Delta\Omega) = 
    \int_{\Delta\Omega} d\Omega\int_{l.o.s.}\rho_{\rm DM} [r(s)]ds \,.
\end{eqnarray}
The radial coordinate $r$ from the center of the halo is expressed as $r = \big(s^2 +r_{\odot}^2-2\,r_{\odot}\,s\, \cos\theta \big)^{1/2}$, where $r_\odot$ is the distance of the observer to the GC, taken to be $r_\odot$ = 8.3 kpc, and $\theta$ corresponding to the angle between the direction of observation and the GC. In the next subsection, we will discuss in detail the distinct  DM density profiles considered in our analysis. However, it is important to emphasize that they will be normalized to the same local DM density, such that $\rho$($r_\odot$ = 8.3 kpc) = 0.38 GeV cm$^{-3}$. This value is derived 
from LAMOST DR5 and Gaia DR2~\cite{2020MNRAS.495.4828G}. As a consequence, any change in the local density can be taken into account by  rescaling the expected signal by ($\rho$($r_\odot$)/0.38 GeV cm$^{-3}$).

In Ref.~\cite{HESS:2016pst}, the authors showed the existence of a CR proton accelerator in the inner 10 pc of the GC, with a harder flux ($\Gamma \simeq$ 2.4) compared to the flux measured in the local Solar neighborhood. This results in a TeV CR proton density higher than that of the local one~\cite{HESS:2016pst,Gaggero:2017jts}. The proton flux normalization in regions around the GC is chosen such that the proton density matches the measured averaged density by H.E.S.S. in the inner 200 pc of the MW, \textit{i.e}, $\omega_{\rm p}^{\rm GC}$($\ge$10 TeV) = 6$\times 10^{-3}$ eV\,cm$^{-3}$, such as
\begin{equation}
    \begin{aligned}
    w_{\rm p}^{\rm GC}(E_{\rm p}\geq10\,\rm TeV) &=  \frac{4\pi}{c} \int_{\rm 10\,TeV}^{\infty} \frac{d\Phi_{\rm p}}{dE_{\rm p}}(E_{\rm p}) \times E_{\rm p}\times dE_{\rm p} \, .
    \end{aligned}
\end{equation}
This is a factor of about six higher than the local proton density $w_{\rm p}^{\rm local}$ derived from proton fluxes measured by AMS~\cite{Consolandi:2016fhd} or DAMPE~\cite{DAMPE:2023pjt}.

Diffusive $\gamma$-ray observations from current and future very-high-energy gamma-ray observatories,  such as H.E.S.S., LHAASO, MAGIC, VERITAS, CTA and SWGO, can be used to place strong upper limits on the inelastic cross-section. One of our goals is to derive  model-independent upper  bounds on $\sigma^{\rm inel}_{\rm p \chi}$.   Although we will not assume a specific model to describe the CR proton-DM interaction, we will  investigate how our findings are dependent on the assumption for the energy dependence of  $\sigma^{\rm inel}_{\rm p \chi} (\sqrt{s})$. Our baseline results will be derived for an energy-independent cross-section above the energy threshold for the pion production in $p \chi$ interactions, \textit{i.e.}, $\sigma^{\rm inel}_{\rm p \chi} (\sqrt{s}) = \sigma^{\rm inel}_{\rm p \chi} (\sqrt{s_0})$, where $\sqrt{s_0}$ is assumed to be 1 GeV in what follows. 
For the energy-dependent case, inspired by Regge theory \cite{Collins:1977jy}, which predicts a power-law behavior for the energy dependence of the proton-proton, 
meson-proton and real photon-proton cross-sections, we will assume 
$  \sigma_{\rm p \chi }^{\rm inel} = \sigma_{\rm p \chi}^{\rm inel}(\sqrt{s_{0}})\times \left(s/s_{0}\right)^{\delta}$. In our analysis, we will consider  two distinct values for the power: $\delta = + 0.096$ and $\delta = - 0.45$, which are the values that describe the current $pp$, $\pi p$, $Kp$ and $\gamma p$ data for high ($\sqrt{s} \gg \sqrt{s_0}$)  and low ($\sqrt{s} \ll \sqrt{s_0}$) energies~\cite{Donnachie:2002en,Menon:2013vka}, respectively. It is important to emphasize that a steeper energy dependence, as observed in the deep inelastic electron/neutrino-proton scatterings \cite{Donnachie:2002en}, could also be considered (see,\textit{e.g.}, Refs. \cite{Soper:2014ska,Guo:2020oum}). Such choice, which is expected to imply stronger bounds, will be analyzed in a forthcoming study, where we will estimate $\sigma_{\rm p \chi }^{\rm inel}$ assuming different models for the $p \chi$ interaction and a detailed description of the photon production in the proton fragmentation. As a consequence, the bounds derived in the current study can be considered as conservative ones.

Finally, inspired by Ref.~\cite{Cyburt:2002uw}, we will investigate if the derived upper bounds for $\sigma^{\rm inel}_{\rm p \chi}$ improve the current exclusion limits on the elastic cross-section obtained in colliders and direct detection experiments. The theoretical description of the connection between these two observables is still an open question, which implies that currently it is only possible to relate the inelastic and elastic cross-sections in a model - dependent way. In our analysis, such a relation will be established assuming that these quantities behave as the corresponding cross-sections for proton - proton interactions. In particular, 
in order to relate the inelastic to the elastic cross-section, we assume either the relation predicted by the \textit{black disc} limit for asymptotic center of mass energies ($\sqrt{s} \rightarrow \infty$) \cite{Collins:1977jy}  such that $\sigma_{\rm p \chi}^{\rm inel} = \sigma_{\rm p \chi}^{\rm el}$, or a relation similar to what is verified in $pp$ collisions at the
LHC energies ($\sqrt{s} \approx 7 - 13$ TeV) ~\cite{Broilo:2020pjr}, $\sigma_{\rm p \chi }^{\rm inel} = 8/3\, \sigma_{\rm p \chi}^{\rm el}$.
Surely, other choices could be considered. However,  
these two choices cover the possibilities that are consistent with the assumption that $\sigma_{\rm p \chi}$ behaves as the $pp$ cross-sections.

\subsection{Dark matter distribution in the Milky Way}
As described in the previous subsection, the expected gamma-ray signal from the CR-DM scattering depends on the assumed DM density distribution. The GC region is an excellent target for DM searches from CR-DM scattering: \textit{(i)} it is a region where we expect high DM density as structure formation indicates that it increases towards the centre of galaxies; and \textit{(ii)} the CR density is found to be significantly
larger in the GeV-to-TeV energies~\cite{Fermi-LAT:2016zaq,HESS:2016pst} than in the local one. The expected signal has a weak dependence on the precise radial DM distribution in the GC region since it scales linearly with the DM density $\rho$ as opposed to the case of annihilating DM where it depends on the square of 
$\rho$. 
Although the highest DM density distribution is expected in the GC region, its precise determination in this complex environment is challenging. Here, we use canonical DM profiles for the description of the DM density distribution in the Milky Way. The baseline DM density distribution chosen is parametrized by the Navarro–Frenk–White (NFW)~\cite{Navarro:1995iw} profile given by
\begin{equation}
    \rho_{\rm NFW}(r)= \frac{\rho_{s}}{(r/r_s)(1+r/r_s)^2}
    \label{eq:NFW}
\end{equation}
with $\rho_{s}$ 
and $r_{s}$ 
 the scale density and radius, respectively. Alternative parametrisations are given by the Einasto~\cite{1965TrAlm...5...87E}  and Burkert~\cite{Burkert:1995yz} density profiles, written as 
\begin{equation}
  \rho_{\rm Ein}(r) = \rho_{s} \exp \left\{-\frac{2}{\alpha}\left[\left(\frac{r}{r_{s}}\right)^{\alpha} - 1\right]  \right\} \quad {\rm and} \quad  
   \rho_{\rm Bur}(r)= \frac{\rho_{s}}{(1 + r/r_s)(1 + (r/r_s)^2)}
   \, ,
    \label{eq:alternative}
\end{equation}
respectively, with $\alpha$ = 0.17 being a sharpness index. The values of the parameters for each density profile are given in Tab.~\ref{tab:profiles}.
\begin{table}[t]
    \centering
    \begin{tabular}{c|c|c|c|c}
        \hline
           \hline
         & Einasto~\cite{HESS:2022ygk} & NFW~\cite{Montanari:2022buj} & cNFW~\cite{Montanari:2022buj} & Burkert~\cite{Cirelli:2010xx}\\
         \hline
        $\rho_{\rm s}$ (GeVcm$^{-3}$) & 0.077 & 0.50 & 0.24 & 0.93\\
        \hline
        $r_{\rm s}$ (kpc) & 20.0 & 15.5 & 23.8 & 12.6\\
        \hline
           \hline
    \end{tabular}
    \caption{DM profile parameters used in this work. Scale density $\rho_{\rm s}$ and radius $r_{\rm s}$ values are given for the Einasto, NFW, cNFW and Burkert profile parametrizations. 
    All the profiles are normalized such that $\rho$(r= r$_\odot$) = 0.38 GeVcm$^{-3}$ assuming r$_\odot$ = 8.3 kpc~\cite{2020MNRAS.495.4828G}.}
    \label{tab:profiles}
\end{table}

In addition to these canonical profiles, we make use a profile parametrization derived from a determination of the MW mass profile using Gaia DR2 measurements
of the rotation curve together with a modeling of
the baryonic components in the GC~\cite{Cautun:2019eaf}.  
The profile is contracted compared to the standard NFW profile by the presence of baryons. The method followed in Ref.~\cite{Montanari:2022buj} does not allow for a precise determination within the inner kpc of the MW, and we will conservatively assume a cored density with 1 kpc~\cite{Montanari:2022buj}. Such a profile is hereafter referred to as the cNFW profile.
However, the DM density could still increase towards the inner GC as shown in several realizations of the DM distribution of MW-like galaxies from realistic  
N-body cosmological simulations including baryonic effects~\cite{Hopkins:2017ycn,McKeown:2021sob}.
We extracted a representative DM distribution from Ref.~\cite{McKeown:2021sob},
where various hydrodynamical simulations using
FIRE-2 zoom-in simulations on MW-like galaxies are performed including 
radiative heating and cooling of gas, stellar feedback
from OB stars, type Ia and type II supernovae, radiation
pressure, and star formation effects. A DM profile for representative DM-only FIRE-2 simulation is also extracted. 
\begin{figure}[h!]
    \centering
    \includegraphics[scale = 0.8]{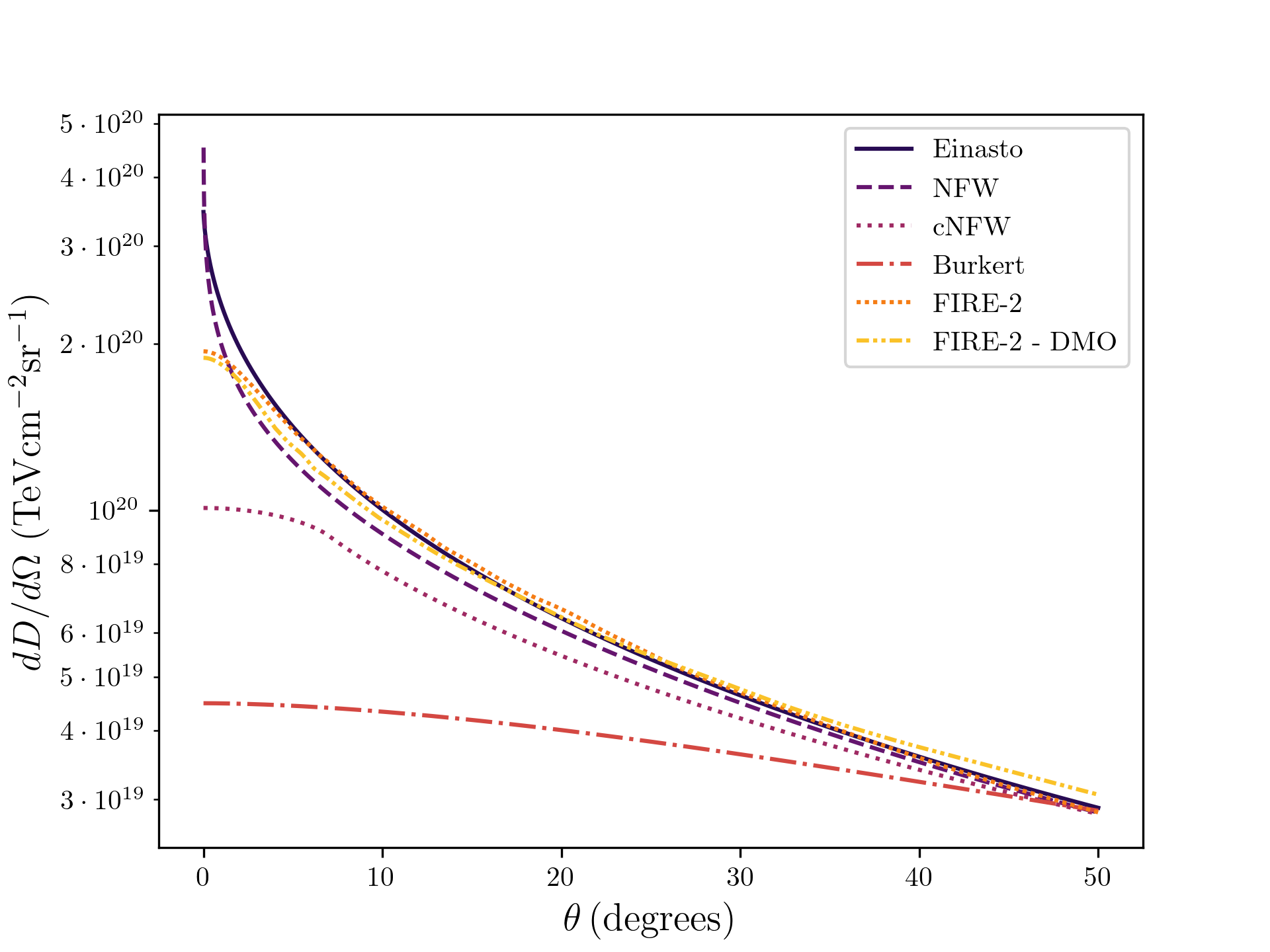}
    \caption{Differential D-factor profiles, expressed in TeV cm$^{-2}$sr$^{-1}$ versus the angular distance from the GC, $\theta$. Various DM profiles are considered: Einasto (dark blue solid line), NFW (purple dashed line), cNFW (burgundy dotted line), Burkert (red dotted-dashed line) as well as those obtained from FIRE-2 cosmological simulations with (orange dotted line) and without (yellow dash-double-dotted line) baryonic feedback. See text for more details.}
    \label{fig:Dfactor}
\end{figure}

Figure~\ref{fig:Dfactor} shows the differential D-factors versus angular distance from the GC for the selected benchmark DM distributions in the inner MW halo. When integrated within the inner 3$^\circ$ of the GC, a factor of $\sim$4.4 is expected between the D-factor, and hence the signal expectation, computed for the two extreme cases considered here, \textit{i.e.}, the Einasto and Burkert profiles.

\subsection{VHE gamma-ray backgrounds in the regions of interest}
\begin{figure}[h!]
    \centering
    \includegraphics[scale = 0.8]{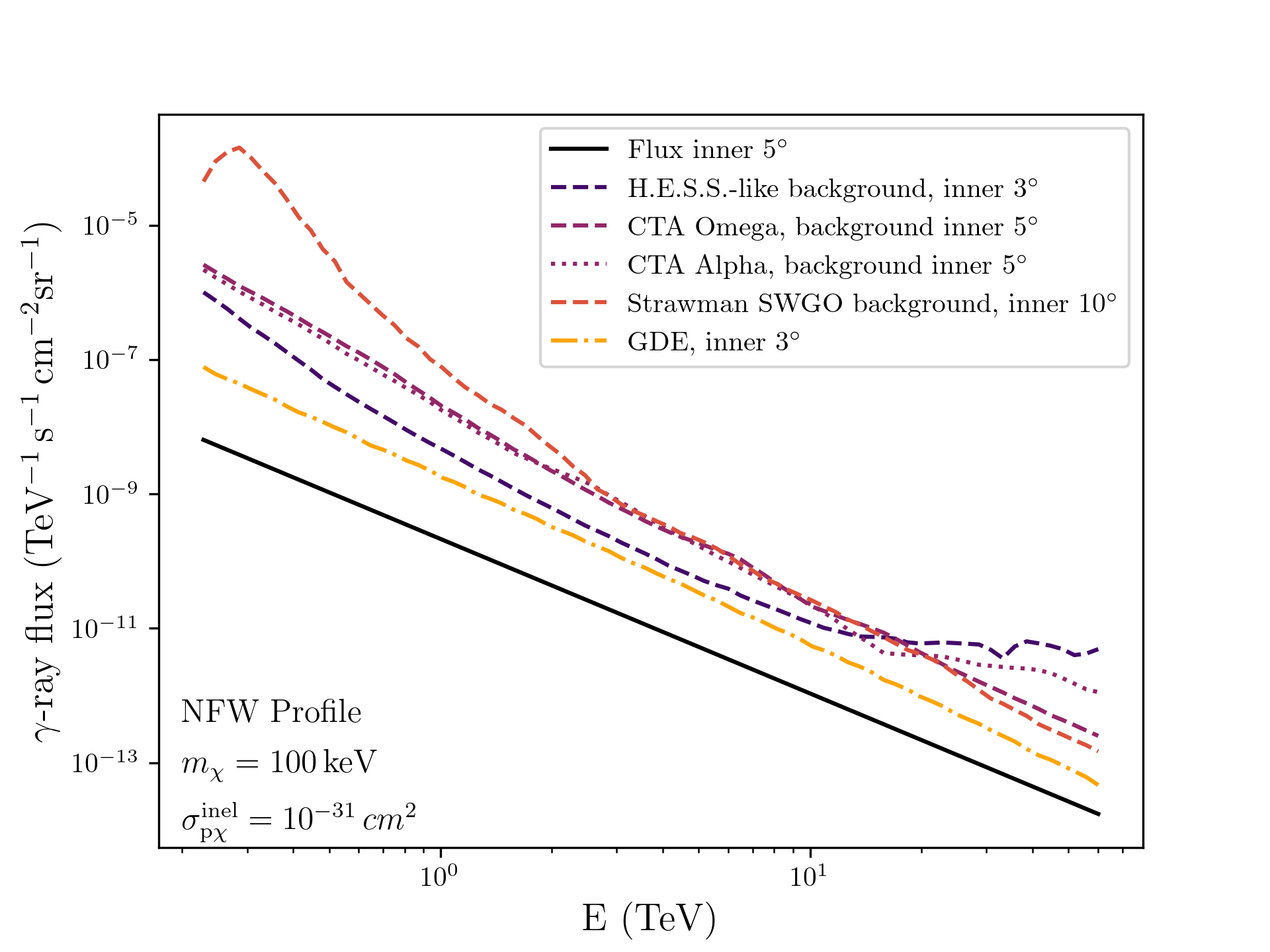}
    \caption{Energy-differential gamma-ray fluxes expected for the DM-induced signal (solid line) 
    in the inner 5$^\circ$ of the GC region.
    The energy-differential fluxes of the residual background (dashed lines) are plotted for H.E.S.S.-like (dark blue lines), CTA (purple lines) and SWGO (red lines) ROIs, respectively. For SWGO, strawman performances are extracted from Ref.~\cite{SWGOirf}.
    The expected signals  
    assume a DM density distribution following the NFW profile populated by DM particles of mass $m_{\rm \chi}$ =  100 keV, and an energy-independent inelastic $p \chi$ cross-section taken to be 
    $\sigma^{\rm inel}_{\rm p \chi}$ = 10$^{-31}$ cm$^{2}$.
    The energy-differential flux of the GDE for the SA0 (R12) model and averaged over the inner 3$^\circ$ is also plotted (dotted-dashed yellow line). See text for further details.} 
    \label{fig:flux}
\end{figure}

An irreducible background arises from the limited discrimination of the instruments considered here between true gamma-ray and cosmic-ray hadrons and electrons, hereafter referred to as the residual background. Its determination 
can be performed via careful Monte Carlo simulations of the response of the instrument or via measurements in blank-field observations in the same observational conditions as for the signal measurement (see Sec.~\ref{sec:observatories} for more details). The energy-differential fluxes per solid angle for the residual background  together with the expected signal are shown in Fig.~\ref{fig:flux} for H.E.S.S.-like observations as well as CTA,  and strawman SWGO expected observations. 
Current-generation IACTs are able to control the systematic uncertainties in the background measurement down to 1\% (see, \textit{e.g.}, Ref.~\cite{HESS:2022ygk}) which allows the detection of sources down to 1\% signal-over-noise ratio provided sufficient statistics can be accumulated.

The central region of the MW is a very rich and complex region of the sky in VHE gamma rays. In addition to the bright sources such as the central source HESS J1745-290~\cite{Aharonian:2009zk,HESS:2016pst}, the supernova remnant HESS J1745-303~\cite{Aharonian:2008gw} as well as various emitters lying close by in the Galactic plane~\cite{HESS:2018pbp}, a faint diffuse emission has been detected, including VHE diffuse emission  attributed to PeV hadrons accelerated in the vicinity of the supermassive
black hole Sgr A*~\cite{HESS:2016pst}. In order to avoid complex modelling of these emissions, we will mask Galactic latitudes $|b|<$0.3$^\circ$ 
in the analysis of the inner GC region for H.E.S.S.-like, CTA and SWGO observations.

Energetic cosmic rays interact with the ambient target materials of the interstellar medium as well as interstellar photon fields, producing a diffuse flux of gamma rays known as the Galactic diffuse emission (GDE). Such gamma rays are produced from a combination of Bremsstrahlung from cosmic-ray hadrons as well as 
the decay of neutral pions produced from these hadrons colliding with the
interstellar gas, and from cosmic-ray electrons scattering off ambient photon fields via inverse Compton scattering (ICS). The GDE contribution dominates the photons seen by Fermi-LAT at MeV-to-GeV energies~\cite{FermiLAT:2012aa}. With the current and expected deep observations of the GC region, GDE is expected to be a relevant background component at TeV energies. In the present work, we will include a realistic model of the expected TeV-scale GDE including the $\pi^0$ and ICS components from the most recent release of the CR propagation framework \texttt{GALPROP} (version 57)~\cite{Porter:2021tlr}. 
We will make use of the SA100 (R12) model as a model for the GDE predictions with \texttt{GALPROP} extracted from Ref.~\cite{Montanari:2023sln}. A detailed study of the GDE impact on the sensitivities to sub-GeV DM is out of the scope of the present paper, however, we will show how much the sensitivity can be improved when the
GDE is subtracted from the measured residual background for the H.E.S.S.-like and CTA observations.
The expected energy-differential fluxes for the GDE including both the $\pi_0$ and ICS components for the SA0 (R12) GDE model is shown in Fig.~\ref{fig:flux}.

\section{Current and near-future VHE gamma-ray observatories}
\label{sec:observatories}
\subsection{H.E.S.S.}
The High Energy Stereoscopic System (H.E.S.S.) consists of an array of five Imaging Atmospheric Cherenkov Telescopes (IACTs) located in the Khomas Highlands of Namibia near the tropic of Capricorn at an altitude of 1800 m~\cite{HESS}. It is the only operating array of IACTs in the Southern hemisphere. H.E.S.S. is sensitive to gamma rays from a few ten GeV to several tens TeV and reaches an energy resolution of $\sigma$/E $\simeq$  10\% and an angular resolution of 0.06$^\circ$ at 68\% containment radius. 

Its location allows the observation of the inner few degrees of the MW under favorable conditions. The GC can be observed from March to September each year with observational zenith angles lower than 30$^\circ$. Observations at low zenith angle are usually preferred since they provide lower energy threshold and control of observational systematic uncertainties, given that atmosphere layers are thinner along the line of sight. However, observations at high zenith angles increase the effective area at higher energies to the price of a raise in the energy threshold
and degraded event reconstruction performances and potentially higher systematic uncertainties (see, for instance, Refs.~\cite{Adams:2021kwm,MAGIC:2022acl} for further discussions).

In order to assess the constraints on DM from CR scattering from current IACTs in the central region of the Milky Way, we will make use of the  the most up-to-date H.E.S.S.-like observations of the inner Galactic halo with the full five-telescope array~\cite{HESS:2022ygk}. These observations are taken thanks to the Inner Galaxy Survey program, which provides the highest exposure in VHE gamma rays so far, for a total of 546 hours of high quality data distributed over a few degrees around the GC. In the following, we consider a region of interest (ROI) of 3$^\circ$ in radius around the GC with a homogeneous time-exposure of 500 hours. The ROI is further divided into 27 concentric rings of 0.1$^\circ$ width. The energy-dependent acceptance and residual background flux for five-telescope H.E.S.S. observations of the inner halo of the Milky Way is extracted from Ref.~\cite{HESS:2022ygk}. We make use of 67 logarithmically-spaced energy bins in the 170 GeV - 67 TeV energy range. This enables us to obtain realistic instrument response for the region of interest we consider. 

\subsection{LHAASO}
The Large High Altitude Air Shower Observatory (LHAASO) is a large air shower observatory currently operating in southwest China~\cite{LHAASO}, set at an altitude of 4,410 meters. It is  composed of the  kilometer square array (KM2A), the water Cherenkov detector array (WCDA), and the wide field of view Cherenkov telescope array (WFCTA). LHAASO detects gamma rays from 100 GeV up to 1 PeV. The performances in terms of acceptance, energy and angular resolutions depend on the considered subarray. In what follows, we will make use of the observations taken by KM2A. It consists of a full-duty extensive air shower array with gamma/hadron separation capability by using both electromagnetic particle detectors and underground muon detectors. At 30 TeV, the angular and energy resolutions of KM2A are about 0.4$^\circ$ (68\% containment radius) and 20\%, respectively, for zenith angles below 20$^\circ$~\cite{Aharonian_2021}.

Given its location in the Northern hemisphere, LHAASO is not suited for observations of the GC region. However, the searched signal is expected to be higher at smaller galactocentric distances. We will therefore consider the region of interest around $15^\circ < b < 45^\circ$ and 30$^\circ < l < 60^\circ$ with a solid angle of 0.274 sr, as described in Ref.~\cite{PhysRevLett.129.261103}. Due to the coordinates of the region of interest, for LHAASO, the local flux of CRs (as measured in the Solar neighborhood) will be used in the signal calculations instead of the flux in the GC region. Four alternative ROIs were considered with the same solid angle and declination for the estimate of the residual background. In what follows, we will assess the expected sensitivity of LHASSO for 340 days of observations with 1/2-KM2A and 230 days with 3/4-KM2A using the energy-dependent acceptance and background rates with 6 logarithmically-spaced energy bins in the 100 TeV -1.5 PeV energy range provided in Ref~\cite{PhysRevLett.129.261103}.

\subsection{CTA}
The Cherenkov Telescope Array (CTA)~\cite{CTA, CTAConsortium:2017dvg} is the next-generation of IACTs with one array in the southern hemisphere at the Paranal site (Chile) and one in the northern hemisphere in La Palma (Spain). It will observe the gamma-ray sky from 20 GeV up to 300 TeV with a factor up-to-ten increase in sensitivity, and angular resolution improved by a factor 2-to-3 compared to currently operating IACTs, and an energy resolution down to
as low as 5\% at TeV energies.
As the GC region is a prime target in the context of CR-DM scattering, 
for this work we will use as baseline the expected performance of the so-called \textit{Omega configuration} array in the Southern hemisphere. This full-scope configuration is expected to be composed of 4 large-sized, 25 medium-sized,
and 70 small-sized telescopes. However, the current initial phase of the CTAO, at Paranal Observatory, will consist of 
14 medium-sized Telescopes and 37 small-sized Telescopes,
known as the Alpha configuration. The performances of these two configurations will be compared.

We will consider a circular ROI of 5$^\circ$ in radius around the GC, excluding Galactic latitudes between $\pm$0.3$^\circ$, in order to avoid complex modeling of the astrophysical sources in the GC plane region, 
with an homogeneous time-exposure of 500 hours. The ROI is further divided into 47 concentric annuli of 0.1$^\circ$ width each.
The Instrument Response Functions (IRF, \textit{i.e.}, energy-dependent effective area, angular and energy resolutions)  computed from Monte Carlo simulations from the Omega configuration array are taken from 
the \textit{publicly available} \texttt{prod3b-v2} IRF library~\cite{CTAOirfprod3}
using 92 logarithmically-spaced energy bins between 50 GeV and 140 TeV.
In particular, the IRF file \texttt{South\_z20\_average\_50h} is used as optimized for the detection of a point-like source at zenith angle of 20$^\circ$. As for the Alpha configuration array, the IRFs ate taken from the \textit{publicly available} \texttt{prod5-v0.1} library~\cite{CTAprod5} using the same 92 logarithmically-spaced energy bins. The file \texttt{Performance-prod5-v0.1-South-20deg} is used as optimized for the detection of a point-like source at zenith angle 20$^{\circ}$.

\subsection{SWGO}
Southern Wide-field Gamma-ray Observatory (SWGO) is the project to build a water Cherenkov particle detector with a wide field of view and  high duty-cycle currently in the research and development phase, and planned to be located in South America at an altitude of 4.4 km or higher and latitude between 10 and 30 degrees south. This gamma-ray observatory is planned to be made of detector units 
which could be deployed as an array or assembled in a building covering an energy range between a few hundred GeV up to the PeV scale~\cite{Conceicao:2023tfb}.

In this work, we use the instrument response functions of the \textit{Strawman} design, as described in the Science Case white paper for the SWGO~\cite{Albert:2019afb}, which are \textit{publicly available}~\cite{SWGOirf}. We consider ten years of observation time with SGWO looking at the inner 10$^\circ$ of the inner MW, excluding Galactic latitudes $|b| < 0.3^\circ$ to assess the SWGO sensitivity. The ROI is further divided into of 49 concentric annuli of 0.2$^\circ$ width each, following the procedure of Ref.~\cite{Viana:2019ucn}. We make use of  46 logarithmically-spaced energy bins in the 200 GeV - 670 TeV energy range.

\section{Statistical analysis framework}
\label{sec:stats}

\subsection{Expected number of signal and background
photons}
For a given density profile, the expected number of signal events, $N^{\rm S}_{\rm ij}$ in the $i$th ROI 
with a solid angle $\Delta\Omega_{\rm i}$,
and $j$th energy bin having a width $\Delta E_{\rm j}$, is given by
\begin{equation}
    N^S_{ij} = T_{{\rm obs}, i} \int_{E_j - \Delta E_j /2}^{E_j + \Delta E_j /2} dE \int_{-\infty}^{\infty} dE'\, \frac{d\Phi^S_{ij}}{dE'} (\Delta \Omega_i, E')\, A_{\rm eff}^{\gamma}(E')\, G(E - E').
\label{eq:ns}
\end{equation}
Here E and E' are the true and reconstructed energies,
respectively, d$\Phi^S_{ij}$/dE' is the predicted gamma-ray flux from CR-DM scattering, as given in
Eq.~(\ref{eq:flux}), T$_{\rm obs,i}$ corresponds to the observation time in the $i$th ROI, A$^\gamma_{\rm eff}$ is the
energy-dependent effective area to gamma rays, and $G$ stands for the finite energy resolution of the instrument which is modeled  as a Gaussian with
width specified by $\sigma$/E.
The number of background events can be obtained straightforwardly from Eq.~(\Ref{eq:ns}) from the substitution of
$d\Phi_{ij}^{\rm S}/dE$ by the energy-differential background flux, $d\Phi_{ij}^{\rm B}/dE$, for both residual and GDE backgrounds.

The IRFs used for each observatory are mentioned in Sec.~\ref{sec:observatories}. They enable us to provide a realistic description of the survey we consider for the computation of the sensitivity. Homogeneous exposure over all the ROIs are considered unless otherwise stated. Given the limited field of view 
of IACTs, a pointing strategy should be defined with a grid of pointing positions in order to reach such an exposure. However, the optimization of the pointing strategy is beyond the scope of this work.  

\subsection{Analysis technique and sensitivity computation}
\label{subsec:stats}
The statistical analysis and the computation of the expected sensitivity of the observatories considered here (see Sec.~\ref{sec:observatories}) are performed following the definition of a log-likelihood ratio test statistic (TS), a well-establish procedure in VHE gamma rays (see, for instance, Refs.~\cite{Abdallah:2016ygi, Abdallah:2018qtu, HESS:2022ygk}). The TS exploits the expected spectral and spatial features of the searched signal. The 2-dimensional binned likelihood function for a given comparison between a theoretical model and observations writes as
\begin{equation}
\mathcal{L}_{ij} (N_{ij}^S, N_{ij}^B, \bar{N}_{ij}^S, \bar{N}_{ij}^B\, |\, N_{ij}^{\rm ON}, N_{ij}^{\rm OFF})     = \textrm{Pois}[(N^S_{ij} + N^B_{ij}), N^{\rm ON}_{ij}] 
    \times \textrm{Pois}[(\bar{N}^S_{ij} + \bar{N}^B_{ij}), N^{\rm OFF}_{ij}] \, ,
\label{eq:likelihood}
\end{equation}
with $\textrm{Pois}[\lambda,n] = e^{-\lambda}\lambda^n/n!$. $N^{\rm ON}_{ij}$ and $N^{\rm OFF}_{ij}$ are the number of measured events in the ON and OFF regions, respectively, in the spectral bin $i$ and the spatial bin $j$.
$N^S_{ij} + N^B_{ij}$ and $\bar{N}^S_{ij} + \bar{N}^B_{ij}$ denote the  expected numbers of events in the signal and background regions, respectively. The background is assumed to be measured from extragalactic blank-field observations or computed from Monte Carlo simulations (see Sec.~\ref{sec:observatories}) such that $\bar{N}^S_{ij}$ can be considered negligible compared to $N^S_{ij}$. 

The full likelihood function is given by the product of the likelihood functions (Eq.~(\ref{eq:likelihood})) over all spatial and energy bins.
The remaining nuisance parameter is the background rate, and it is estimated by maximization of the profile
likelihood. Therefore, the likelihood will depend only on the DM particle properties, which are parameterized here by $\sigma^{\rm inel}_{\rm p\chi}$ and m$_{\chi}$. For a given DM mass m$_{\chi}$, the likelihood is then used to define the log-likelihood ratio test statistic (TS) for setting upper limits, such as
\begin{equation}
    \text{TS}(m_{\chi}) = - 2 \ln \frac{\mathcal{L}(\sigma^{\rm inel}_{\rm p\chi}, m_{\chi})}{\mathcal{L}(\widehat{\sigma^{\rm inel}_{\rm p\chi}}, m_{\chi})} \, ,
    \label{eq:TS}
\end{equation}
where $\widehat{\sigma^{\rm inel}_{\rm p\chi}}$ corresponds to the value of the inelastic cross-section which maximizes the likelihood for a given $m_{\chi}$.
In the large-statistics limit, this TS follows a $\chi^2$ distribution with a one degree of freedom. Assuming this limit holds, we can set one-sided 95\% upper limits on $\sigma^{\rm inel}_{\rm p\chi}$ by solving Eq.~(\ref{eq:TS}) for the cross-section above the best fit, where
TS = 2.71. This is the procedure we use in the following to compute our limits.

\begin{figure}
    \centering
    \includegraphics[width = 0.8\textwidth]{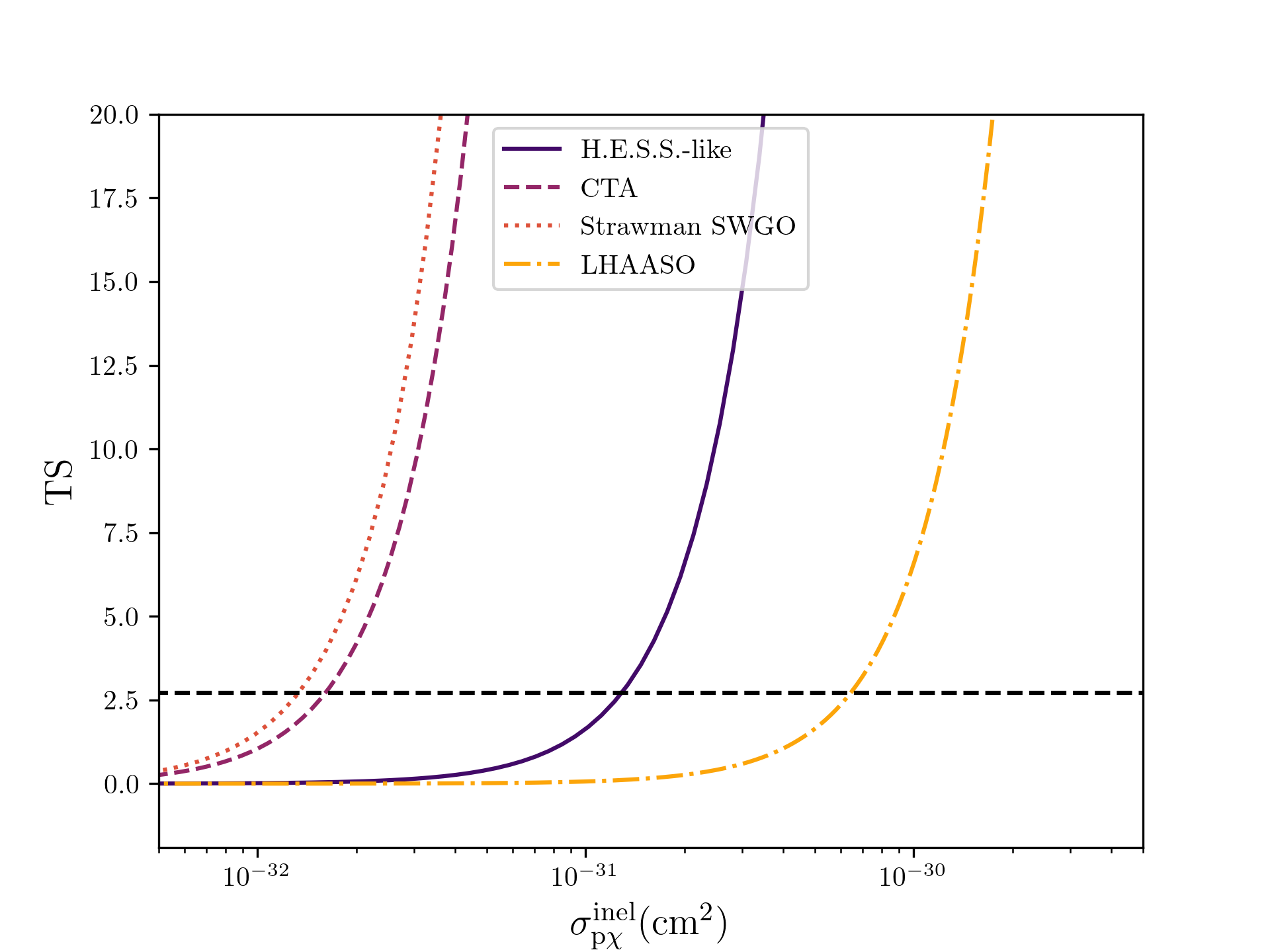}
    \caption{Log-likelihood ratio TS profile curves for
    H.E.S.S.-like (dark blue solid line), CTA (purple dashed line), strawman SWGO (orange dotted line) and LHAASO (yellow dotted-dashed line)  observations
    as a function of the inelastic cross-section $\sigma^{\rm inel}_{\rm p \chi}$ assuming a NFW DM profile and a DM mass of 1 MeV. The black dashed  horizontal line represents TS = 2.71 used to derive  95\% C.L. value on $\sigma^{\rm inel}_{\rm p \chi}$.}
    \label{fig:TS}
\end{figure}
The mean expected upper limits and confidence intervals can be computed by applying the above-mentioned procedure to a large number of Monte Carlo realizations of mock datasets of $N^{\rm ON}_{ij}$ and $N^{\rm OFF}_{ij}$ and analyzing the distribution of the obtained limits. This procedure is largely used in IACT analyses, and in particular in H.E.S.S.~\cite{Abdallah:2016ygi,Abdallah:2018qtu,HESS:2022ygk}. However, in order to avoid the computational intensive generation of datasets, one can take the mean expected background as a single `representation' of the dataset, known as the \textit{Asimov dataset}~\cite{Cowan:2011an}, in order to compute the mean expected sensitivity. The Asimov procedure can be also used to compute the  $N$-sigma containment interval~\cite{Cowan:2011an}
from ${\rm TS} = (\Phi^{-1}[0.95] \pm N)^2$, where $\Phi^{-1}$ is the inverse of the cumulative distribution function of a normal distribution with mean $\mu=0$ and standard deviation $\sigma=1$.
We will use the Asimov procedure throughout this work for the derivation of the mean expected limits together with their 1$\sigma$ and 2$\sigma$ containment bands.
Figure~\ref{fig:TS} shows the TS profile obtained for the different observatories as a function of the free parameter $\sigma^{\rm inel}_{\rm p\chi}$ for a given DM mass of 1 MeV and a NFW DM density distribution profile for the MW.
\begin{figure}
    \centering
    \includegraphics[width = 0.8\textwidth]{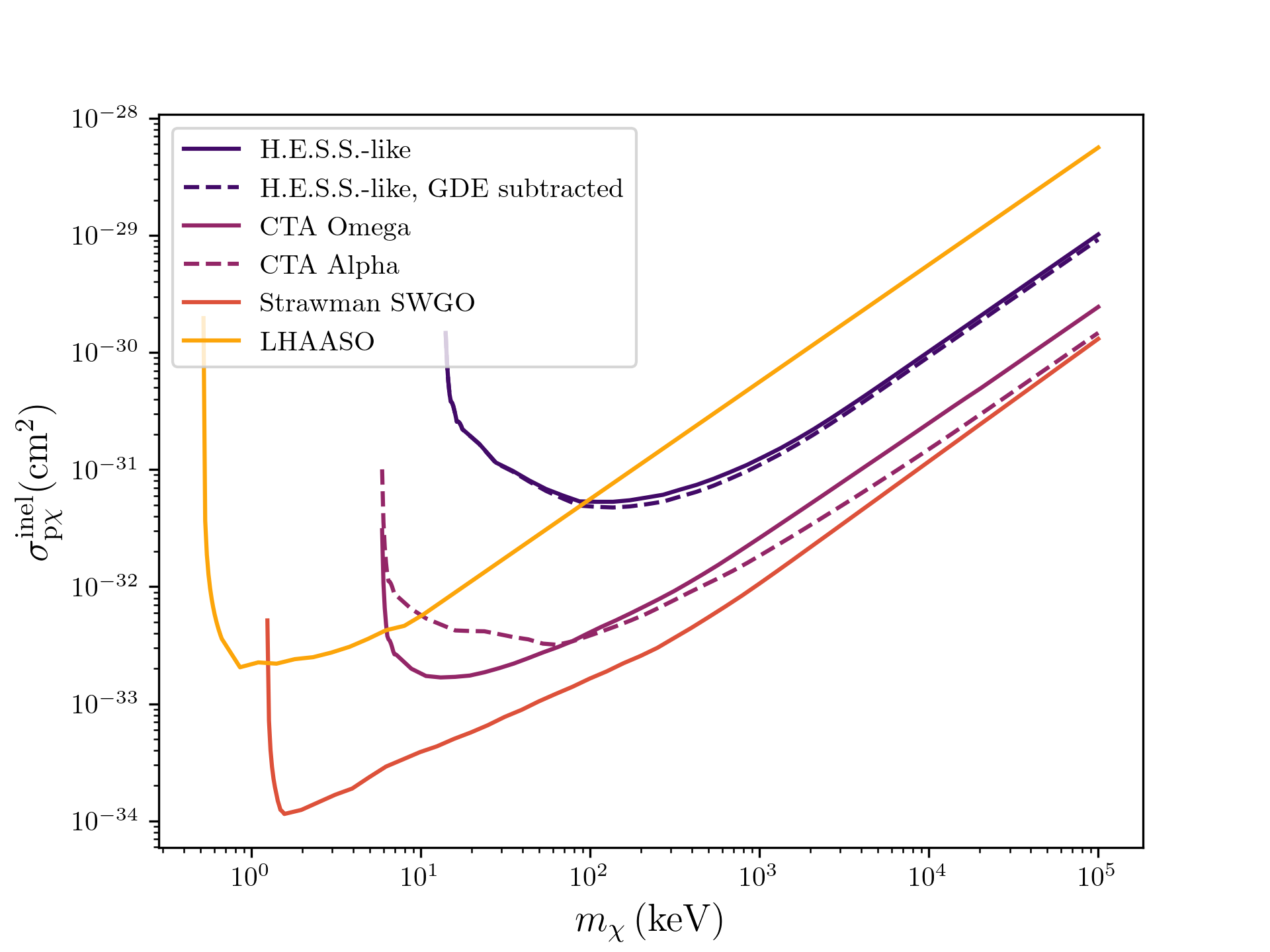}
    \caption{Sensitivity curves for H.E.S.S.-like (blue line), CTA (purple lines), strawman SWGO (orange line) and LHAASO (yellow line)  observations on inelastic scattering of CR protons off DM. The solid and dashed purple lines corresponds to the sensitivity expected for the Omega (baseline) and Alpha configurations, respectively.
    The sensitivity is expressed as 95\% C.L. mean expected upper limit. Here, we assumed a NFW DM profile and an energy-independent inelastic cross-section normalized at $\sqrt{s}$ = 1 GeV. The sensitivity reach for H.E.S.S.-like observations when the GDE component is subtracted is shown as purple dashed line.}
    \label{fig:allexp}
\end{figure}

\section{Results and discussions}
\label{sec:results}
The sensitivity is expressed in terms of mean expected upper limit computed at 95\% C. L. (see Sec.~\ref{subsec:stats}).
We therefore here assess the statistical reach of current and forthcoming VHE observatories\footnote{No systematic uncertainty is assumed in the TS analysis. It should be noted, however, that the systematic uncertainty in the residual background determination are controlled down to 1\% in the H.E.S.S. inner galaxy survey~\cite{HESS:2022ygk}.}.
We choose the NFW profile as the baseline DM density profile and the fraction of the proton energy going into neutral pions to be $f_\pi$ = 0.17,   excepted when explicitly stated otherwise.

Figure~\ref{fig:allexp} shows the sensitivity for H.E.S.S.-like, CTA (Omega and Alpha configurations), SWGO and LHAASO, on the inelastic  cross-section versus DM mass, 
assuming a NFW profile and an energy-independent inelastic cross-section (see Sec.~\ref{subsec:model}). The strongest sensitivities reached for each observatory are the following: for H.E.S.S.-like, $\sigma_{\rm p \chi}^{\rm inel}$ = 5.3$\times$10$^{-32}$  cm$^2$  for a mass $m_{\rm \chi}$ = 108 keV;  for CTA, $\sigma_{\rm p \chi}^{\rm inel}$ = 1.6$\times$10$^{-33}$ cm$^2$ for a mass $m_{\rm \chi}$ = 13 keV; for SWGO, $\sigma_{\rm p \chi}^{\rm inel}$ = 1.0$\times$10$^{-34}$  cm$^2$ for a mass $m_{\rm \chi}$ = 1.4 keV; and for LHAASO, $\sigma_{\rm p \chi}^{\rm inel}$ = 2.0$\times$10$^{-33}$ cm$^2$ for a mass $m_{\rm \chi}$ = 0.8 keV. 
The expected sensitivity from CTA Alpha configuration (purple dashed line) shows a reduced sensitivity by a factor up to 3 in the 10 - 100 keV mass range compared to the Omega configuration sensitivity. 
Above 100 keV masses, the Alpha CTA configuration is about 50\% more sensitive than the baseline configuration due to an increased ratio between signal and square root of the expected background.
In order to understand why the minimum values of $m_{\chi}$ reached for these observatories are different, it is important to remember that the gamma-ray production only occurs if the CR proton has enough energy to produce (at least) one neutral pion in the interaction with a DM of mass $m_{\chi}$. Such threshold defines the value of $E_{\rm p}^{\rm min}$, which is inversely proportional to $m_{\chi}$, and also $E_{\gamma}^{\rm min}$  [see discussion after Eq. (\ref{eq:flux})]. As the accessible gamma-ray energy ranges of the experiments are distinct, it implies that they probe different minimum values of $m_{\chi}$.
In particular, since LHAASO is capable of detecting the highest-energy gamma rays, it can probe the smallest DM masses. However, its sensitivity above 100 keV masses is lower than that of the other experiments due to the smaller CR flux in its ROIs compared to the flux in the GC region. This decrease in sensitivity occurs despite the fact that the observed region and effective area are larger compared to other experiments.
When subtracting the GDE component to the overall background, the H.E.S.S.-like sensitivity improves by 10\%. A slight dependency is obtained with the
DM mass. The residual background becomes significantly larger than the GDE component at the highest energies, which therefore implies that the GDE impact becomes negligible at the lowest masses.

\begin{figure}
    \centering
    \includegraphics[width = 0.8\textwidth]{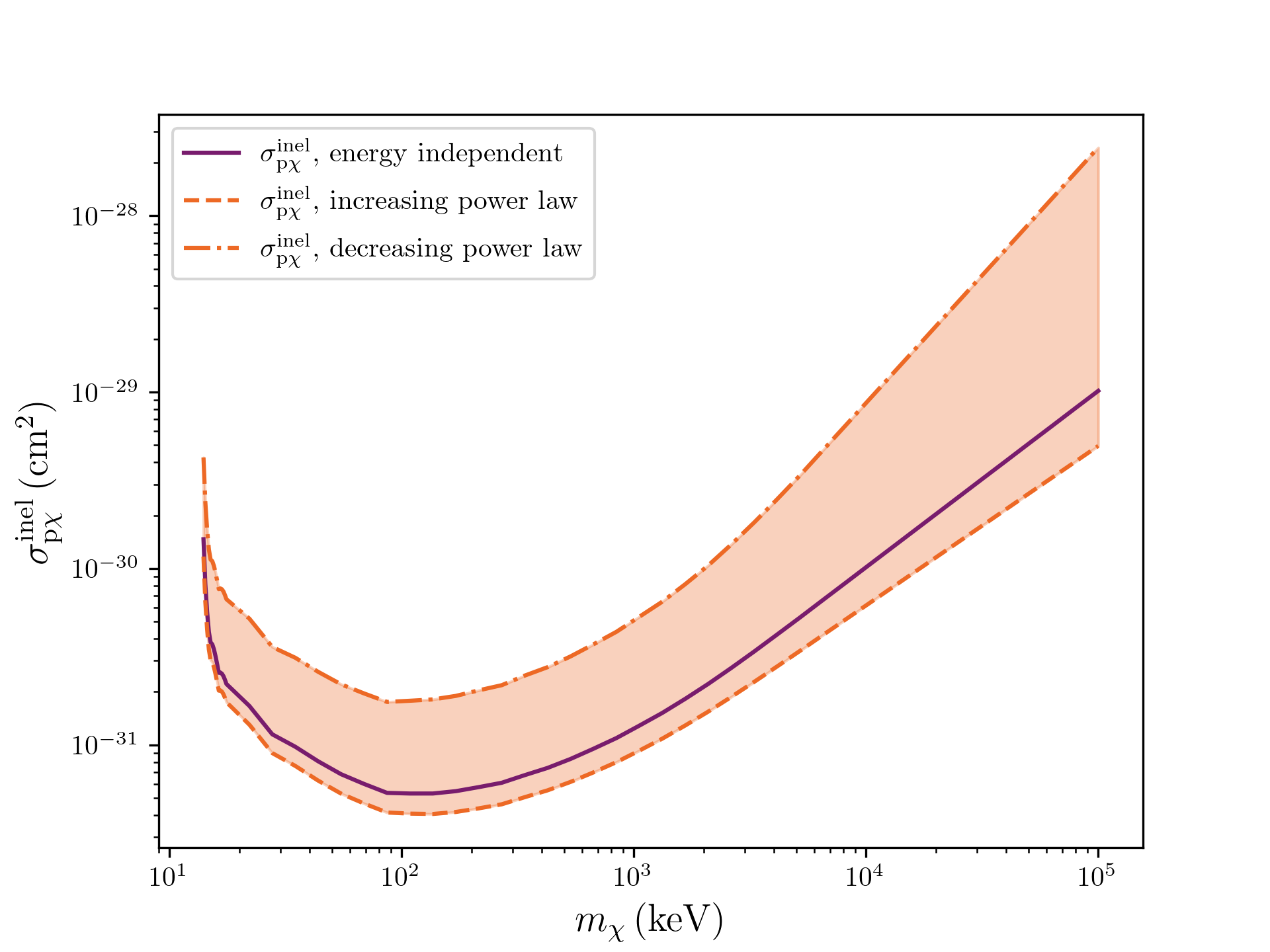}
    \caption{Comparison between the sensitivity reached for H.E.S.S.-like observations for the energy-dependent and independent inelastic cross sections $\sigma^{\rm inel}_{\rm p \chi}$, respectively. The sensitivity is shown as a solid purple line for the constant cross-section, and a band between the energy-dependent power law cases for the inelastic cross-section with spectral indexes of 0.096 (increasing power law, orange dashed line) and -0.45 (decreasing power law, dotted-dashed orange line), respectively. At the lowest masses, it implies a difference of 60\% while a factor of up to $\sim 50$ is obtained at the highest masses.}
    \label{fig:econstrel}
\end{figure}

Figure~\ref{fig:econstrel} shows the effect of the energy dependency for the inelastic $p\chi$ cross-section on the sensitivity curve for H.E.S.S.-like observations. The solid line represents the energy-independent inelastic cross-section between DM and CR protons, respectively. For the energy-dependent case,  
the dashed and dotted-dashed lines correspond to $\propto E^{0.096}$ and $\propto E^{-0.45}$ cases, respectively.
The energy dependence of the inelastic cross-section induces a specific behavior of the mass-dependency of the sensitivity, reaching a factor up to $\sim$ 50 for a DM mass of 100 MeV.
In the energy-independent case, the sensitivity scales approximately with $m_\chi$ in the high-mass limit. 
\begin{figure}
\centering
  \includegraphics[width = 0.49\textwidth]{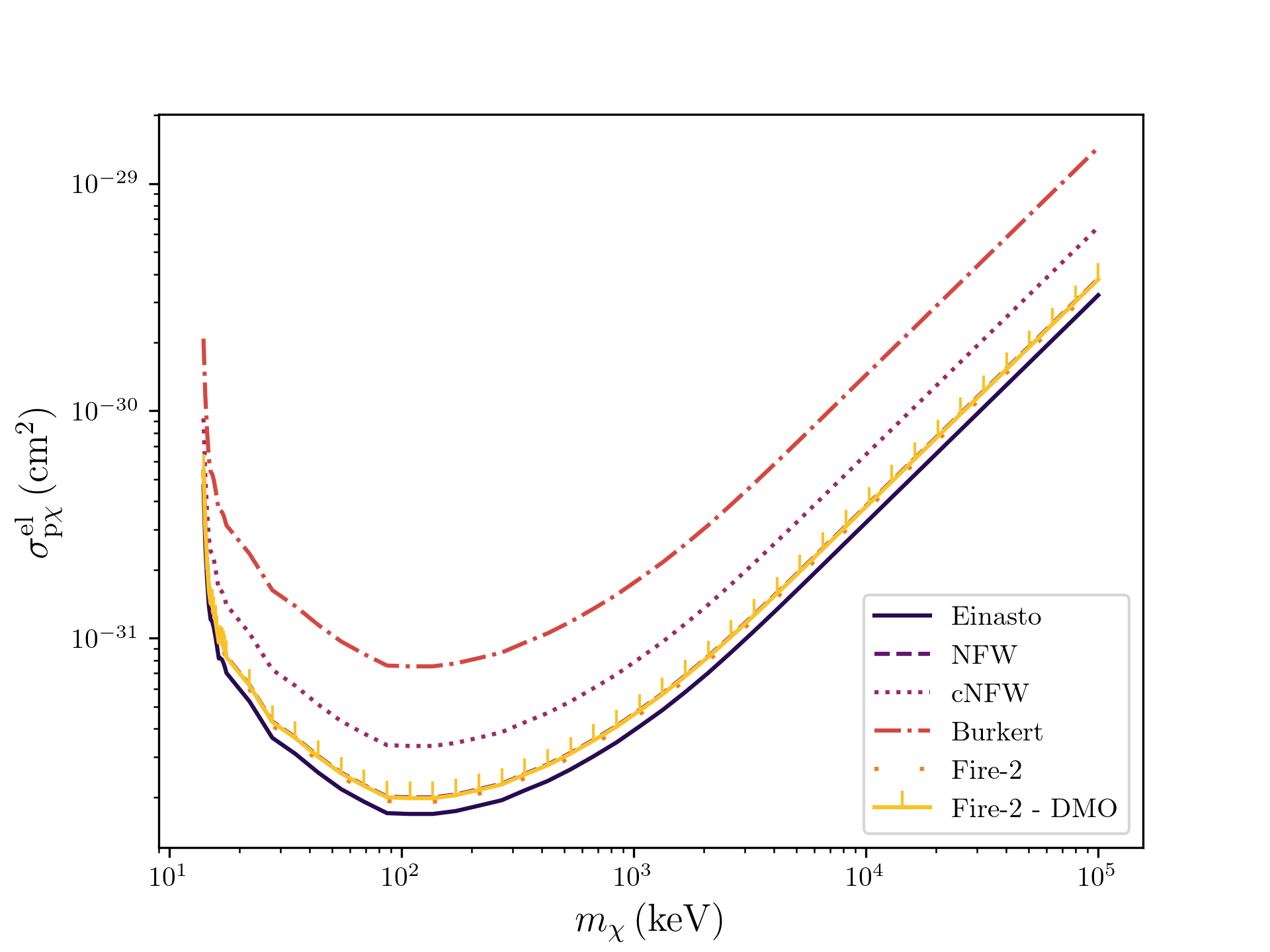}
  \includegraphics[width = 0.49\textwidth]{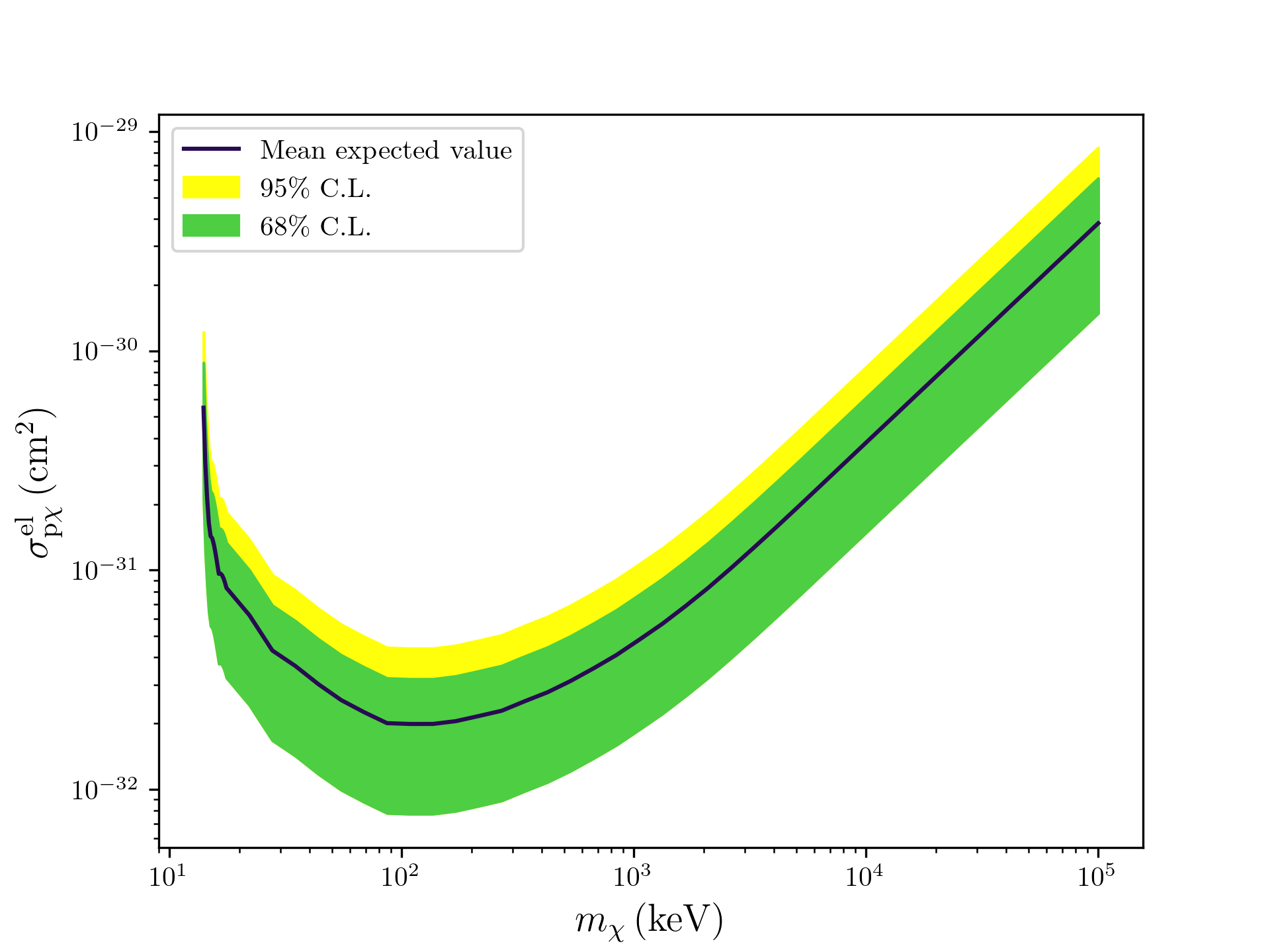}
\caption{ H.E.S.S.-like sensitivity expressed as 95\% C.L. mean upper limit versus DM mass for the elastic CR - DM scattering process. \textit{Left panel:}  
H.E.S.S.-like sensitivity expressed in terms of 95\% C.L. mean expected upper limit for different DM density profiles considered here: Einasto (dark blue solid line), NFW (purple dashed line), cNFW (burgundy dotted line), Burkert (red dotted-dashed line), Fire-2 (orange loosely dotted line) and Fire-2 - DMO (yellow up-ticked line), respectively. The difference between the strongest (Einasto profile) and weakest (Burkert profile) curves is given by a multiplicative factor of $\sim4$. \textit{Right panel:} 
The sensitivity expressed as the mean expected upper limit computed at 95\% C.L. (dark blue solid line) is shown for the H.E.S.S.-like observations together with the included 68\%  (green area) and 95\% (yellow area) confidence level bands, respectively, assuming the NFW DM profile,}
\label{fig:hess}
\end{figure}

In what follows, we investigate if the constraints derived for $\sigma^{\rm inel}_{\rm p \chi}$ can improve the current limits in the elastic proton - DM cross-section, which are usually derived considering direct detection and cosmological bounds. As explained in 
Sec.~\ref{sec:crdm}, we assume that the relation between these cross-sections can be expressed by 
$\sigma^{\rm inel}_{\rm p \chi} = \kappa \cdot \sigma^{\rm el}_{\rm p \chi}$, with $\kappa = 8/3$ or $\kappa = 1$, motivated by the current results for $pp$ collisions at the LHC and by the black disc limit, respectively. 
The left panel of Fig.~\ref{fig:hess} shows our results for $\sigma^{\rm el}_{\rm p \chi}$ on the H.E.S.S.-like sensitivity, derived for $\kappa = 8/3$, considering different DM density profiles. A factor of about 4 is obtained between the strongest sensitivity obtained for the Einasto profile, and the weakest one for the Burkert profile. The right panel of Fig.~\ref{fig:hess}  shows the sensitivity expressed in terms of mean expected upper limits computed at 95\% C. L. together with the  68\% and 95\% confidence level statistical containment bands. We have verified that these results are modified  by a factor  2.6 if the value $\kappa = 1$ is considered in the calculations. Moreover, we also have investigated the impact of a smaller value of $f_{\pi}$, which can occur with the increasing of the neutral pion multiplicity at high energies. Assuming a decrease of $f_{\pi}$ by a factor 3 would degrade the sensitivity by a factor of $\sim$4. Decreasing $f_{\rm \pi}$ would modify the accessible mass range
to higher masses. 
We have verified that the dependence of the expected gamma-ray flux on $\kappa$ and $f_{\pi}$ translates into a straightforward scaling of the sensitivity [see Eq.~(\ref{eq:flux})]. A change by a factor of about 9 is obtained between the scenario with $\kappa$ = 8/3 and $f_{\pi}$ = 0.17, and the scenario with $\kappa$ = 1 and $f_{\pi}$ = 0.17/3 at larger masses. Surely, the uncertainty on these predictions can be reduced by assuming a more detailed description of the CR proton-DM interaction and for the photon production in the final state, which we plan to perform in a forthcoming study.

\begin{figure}
    \centering
    \includegraphics{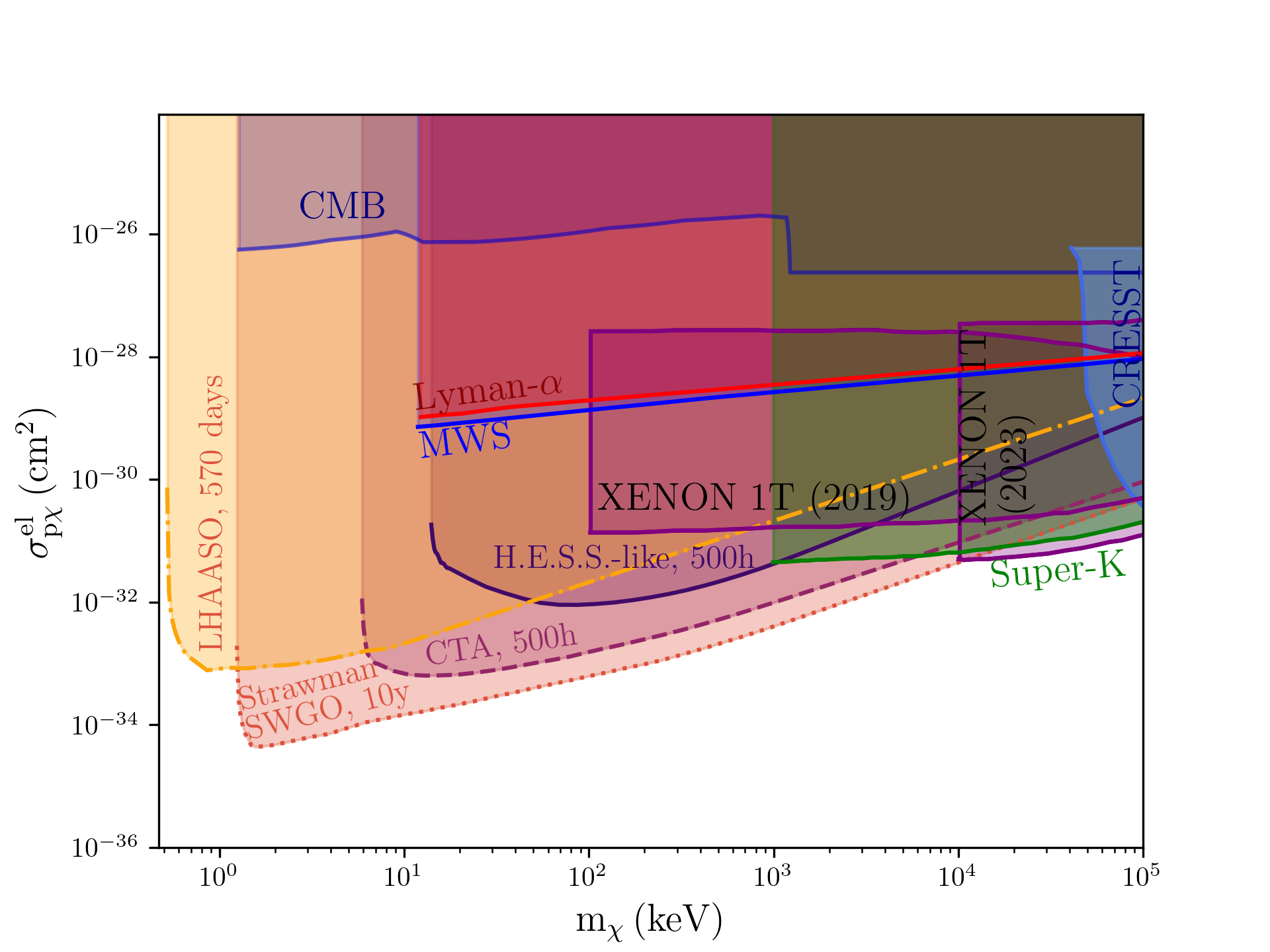}
    \caption{Current constraints on the elastic cross-section between DM and protons. VHE gamma-ray constraints, assuming no energy dependence of the elastic cross-section (see Sec.~\ref{subsec:model}) obtained in the present work are shown for current (H.E.S.S-like and LHAASO) and future (CTA Omega, strawman SWGO) VHE gamma-ray observatories.
    Direct detection experiment constraints derived by CRESST~\cite{Collar:2018ydf} and Xenon1T~\cite{Bringmann:2018cvk, Alvey:2022pad}, neutrino searches with Super-K~\cite{Super-Kamiokande:2022ncz},   and colliders~\cite{Daci:2015hca} searches are shown as well as from cosmological probes such, as CMB~\cite{Gluscevic:2017ywp}, Lyman-$\alpha$~\cite{Murgia:2017lwo,Xu:2018efh,Buen-Abad:2021mvc}, and Milky-Way subhalos (MWS)~\cite{Nadler:2019zrb,Slatyer:2018aqg, Buen-Abad:2021mvc}.}
    \label{fig:limcomp}
\end{figure}

In order to put these results into a wider context and compare them with direct detection, colliders and cosmological constraints, we will use the 
energy-independent case for the inelastic cross-section as well as the relationship between the inelastic and elastic  cross-sections, discussed in Sec.~\ref{subsec:model}, given by a factor 8/3. Figure~\ref{fig:limcomp} shows the sensitivity for sub-GeV DM computed for the observatories considered here together with the excluded regions from other probes such as Cosmological, using CMB~\cite{Gluscevic:2017ywp,Xu:2018efh,Slatyer:2018aqg}, Lyman-$\alpha$~\cite{Buen-Abad:2021mvc} and Milk Way subhalo (MWS)~\cite{Buen-Abad:2021mvc} measurements, 
and direct detection~\cite{Collar:2018ydf, Bringmann:2018cvk, Super-Kamiokande:2022ncz} searches. 
The sensitivity reached for each experiment goes as follows: for H.E.S.S.-like, $\sigma_{\rm p \chi}^{\rm el}$ = 1.9$\times$10$^{-32}$ cm$^{2}$ for a mass $m_{\rm \chi} = $ 108 keV; for CTA, $\sigma_{\rm p \chi}^{\rm el}$ = 6.3$\times$10$^{-34}$ cm$^{2}$ for a mass $m_{\rm\chi} = $ 13 keV; for a strawman SWGO model, $\sigma_{\rm p \chi}^{\rm el}$ = 3.9$\times$10$^{-32}$ cm$^{2}$ for a mass $m_{\rm\chi} = $ 1.4 keV; for LHAASO, $\sigma_{\rm p \chi}^{\rm el}$ = 7.6$\times$10$^{-34}$ cm$^{2}$ for a mass $m_{\rm\chi} = $ 0.8 keV.
This highlights the potential of this novel approach to probe a region of the parameter space currently unavailable to the more traditional collider and direct detection searches. Using a set of conservative values for  $f_{\pi}$ and $\kappa$ would degrade the sensitivity up to one order of magnitude, leaving this method well positioned to cover uncharted regions of the parameter space.
We forecast here the sensitivity of current on ongoing VHE observatories in a model-independent way using approximate relationship between the inelastic and elastic cross-section.
A more accurate comparison would require the definition of an underlying model for the dark matter nature and properties such as coupling strength and mediator mass.

\section{Summary}
\label{sec:sum}
In this work, we have derived predictions for the constraints on the inelastic cross-section between CR protons and DM particles populating the DM halo of the Milky Way based on the latest H.E.S.S. observations of the inner 3$^\circ$ of the MW, and LHAASO observations in the region of the sky defined by 15$^\circ <$ b $<$ 45$^\circ$ and 30$^\circ <$ l $<$ 60$^\circ$ using the PeV CR proton source located in the GC and realistic DM distributions in the Milky Way. The impact of our modelling assumptions is investigated considering 
a variety of viable DM profiles and distinct assumptions for the energy dependence of the inelastic $p\chi$ cross-section. Our results indicate that  VHE gamma-ray observations enable to probe a DM mass range poorly covered so far and provide model - independent upper bounds on the value of $\sigma^{\rm inel}_{\rm p \chi}$.
The strongest limits come from SWGO in the few keV range, reaching an inelastic cross-section of 1.4$\times$10$^{-34}$ cm$^{2}$ for a DM mass of 1.5 keV. The sensitivity for H.E.S.S.-like observations probe inelastic cross-sections down to $5.3\times10^{-32}$ cm$^{2}$ for a 108 keV DM mass.
We also have forecast the sensitivity reach of the future observations of CTA in the central region of the Milky Way, which demonstrates the potential of these observatories. 

The constraints derived on the inelastic cross-section $\sigma^{\rm inel}_{p\chi}$, assuming no specific underlying DM model, have been recast in terms of the elastic cross-section $\sigma^{\rm el}_{\chi p}$ and upper bounds for this quantity have been presented considering the distinct VHE gamma-ray observatories. Moreover, our results have been compared  with other probes such as cosmological, collider and direct detection searches,  which has allowed us to highlight further synergies of the present approach. These results indicate that current and future VHE gamma-ray observations open a new window for constraining inelastic and elastic cross-sections in the poorly charted DM mass range from a hundred eV to a hundred MeV. Further detailed comparisons between the various probes in this DM mass range will require the choice of a specific DM model, which will be carried on in a future work, for the derivation of an accurate relationship between inelastic and elastic cross-section such as couplings, mediator mass, and properties of the DM particles.

\acknowledgments
We warmly thank Alessandro Montanari
for providing GDE emission fits data of Figures 3 and 4 of Ref.~\cite{Montanari:2023sln} for the SA10 (R12) model.
This work was conducted in the context of the CTA Consortium\footnote{\url{https://www.cta-observatory.org/consortium_acknowledgments/}} and the SWGO Collaboration. This research has made use of the CTA instrument response functions provided by the CTA Consortium and Observatory, see
https://www.cta-observatory.org/science/cta-performance for more details,
and the SWGO instrument response functions provided by the SWGO Consortium (see~\cite{SWGOirf}).
This work is supported by the "ADI 2021" project funded by the IDEX Paris ANR-11-IDEX-0003-02, and by FAPESP, process numbers 2019/14893-3, 2021/02027-0 and 2021/01089-1. A.V. is supported by CNPq grant 314955/2021-6. 
V.P.G. was partially supported by CNPq, FAPERGS and INCT-FNA (Process No. 464898/2014-5). A.V. and I.R. acknowledge the National Laboratory for Scientific Computing (LNCC/MCTI,  Brazil) for providing HPC resources of the SDumont supercomputer (http://sdumont.lncc.br).

\bibliographystyle{JHEP}
\bibliography{biblio}

\providecommand{\href}[2]{#2}\begingroup\raggedright\begin{thebibliography}{10}

\bibitem{Kolb:1990vq}
E.W.~Kolb and M.S.~Turner, \emph{{The Early Universe}}, vol.~69 (1990), \href{https://doi.org/10.1201/9780429492860}{10.1201/9780429492860}.

\bibitem{Clowe:2006eq}
D.~Clowe, M.~Bradac, A.H.~Gonzalez, M.~Markevitch, S.W.~Randall, C.~Jones et~al., \emph{{A direct empirical proof of the existence of dark matter}}, \href{https://doi.org/10.1086/508162}{\emph{Astrophys. J. Lett.} {\bfseries 648} (2006) L109} [\href{https://arxiv.org/abs/astro-ph/0608407}{{\ttfamily astro-ph/0608407}}].

\bibitem{Garrett:2010hd}
K.~Garrett and G.~Duda, \emph{{Dark Matter: A Primer}}, \href{https://doi.org/10.1155/2011/968283}{\emph{Adv. Astron.} {\bfseries 2011} (2011) 968283} [\href{https://arxiv.org/abs/1006.2483}{{\ttfamily 1006.2483}}].

\bibitem{Dodelson:2011qv}
S.~Dodelson, \emph{{The Real Problem with MOND}}, \href{https://doi.org/10.1142/S0218271811020561}{\emph{Int. J. Mod. Phys. D} {\bfseries 20} (2011) 2749} [\href{https://arxiv.org/abs/1112.1320}{{\ttfamily 1112.1320}}].

\bibitem{Kahlhoefer:2017dnp}
F.~Kahlhoefer, \emph{{Review of LHC Dark Matter Searches}}, \href{https://doi.org/10.1142/S0217751X1730006X}{\emph{Int. J. Mod. Phys. A} {\bfseries 32} (2017) 1730006} [\href{https://arxiv.org/abs/1702.02430}{{\ttfamily 1702.02430}}].

\bibitem{Schumann:2019eaa}
M.~Schumann, \emph{{Direct Detection of WIMP Dark Matter: Concepts and Status}}, \href{https://doi.org/10.1088/1361-6471/ab2ea5}{\emph{J. Phys. G} {\bfseries 46} (2019) 103003} [\href{https://arxiv.org/abs/1903.03026}{{\ttfamily 1903.03026}}].

\bibitem{Strigari:2018utn}
L.E.~Strigari, \emph{{Dark matter in dwarf spheroidal galaxies and indirect detection: a review}}, \href{https://doi.org/10.1088/1361-6633/aaae16}{\emph{Rept. Prog. Phys.} {\bfseries 81} (2018) 056901} [\href{https://arxiv.org/abs/1805.05883}{{\ttfamily 1805.05883}}].

\bibitem{Ackermann:2015zua}
{\scshape Fermi-LAT} collaboration, \emph{{Searching for Dark Matter Annihilation from Milky Way Dwarf Spheroidal Galaxies with Six Years of Fermi Large Area Telescope Data}}, \href{https://doi.org/10.1103/PhysRevLett.115.231301}{\emph{Phys. Rev. Lett.} {\bfseries 115} (2015) 231301} [\href{https://arxiv.org/abs/1503.02641}{{\ttfamily 1503.02641}}].

\bibitem{HESS:2022ygk}
{\scshape H.E.S.S.} collaboration, \emph{{Search for Dark Matter Annihilation Signals in the H.E.S.S. Inner Galaxy Survey}}, \href{https://doi.org/10.1103/PhysRevLett.129.111101}{\emph{Phys. Rev. Lett.} {\bfseries 129} (2022) 111101} [\href{https://arxiv.org/abs/2207.10471}{{\ttfamily 2207.10471}}].

\bibitem{XENON:2023cxc}
{\scshape XENON} collaboration, \emph{{First Dark Matter Search with Nuclear Recoils from the XENONnT Experiment}}, \href{https://doi.org/10.1103/PhysRevLett.131.041003}{\emph{Phys. Rev. Lett.} {\bfseries 131} (2023) 041003} [\href{https://arxiv.org/abs/2303.14729}{{\ttfamily 2303.14729}}].

\bibitem{XENON:2021qze}
{\scshape XENON} collaboration, \emph{{Emission of single and few electrons in XENON1T and limits on light dark matter}}, \href{https://doi.org/10.1103/PhysRevD.106.022001}{\emph{Phys. Rev. D} {\bfseries 106} (2022) 022001} [\href{https://arxiv.org/abs/2112.12116}{{\ttfamily 2112.12116}}].

\bibitem{Battaglieri:2017aum}
M.~Battaglieri et~al., \emph{{US Cosmic Visions: New Ideas in Dark Matter 2017: Community Report}},  in \emph{{U.S. Cosmic Visions: New Ideas in Dark Matter}}, 7, 2017 [\href{https://arxiv.org/abs/1707.04591}{{\ttfamily 1707.04591}}].

\bibitem{Ambrosone:2022mvk}
A.~Ambrosone, M.~Chianese, D.F.G.~Fiorillo, A.~Marinelli and G.~Miele, \emph{{Starburst Galactic Nuclei as Light Dark Matter Laboratories}}, \href{https://doi.org/10.1103/PhysRevLett.131.111003}{\emph{Phys. Rev. Lett.} {\bfseries 131} (2023) 111003} [\href{https://arxiv.org/abs/2210.05685}{{\ttfamily 2210.05685}}].

\bibitem{Elor:2021swj}
G.~Elor, R.~McGehee and A.~Pierce, \emph{{Maximizing Direct Detection with Highly Interactive Particle Relic Dark Matter}}, \href{https://doi.org/10.1103/PhysRevLett.130.031803}{\emph{Phys. Rev. Lett.} {\bfseries 130} (2023) 031803} [\href{https://arxiv.org/abs/2112.03920}{{\ttfamily 2112.03920}}].

\bibitem{Bringmann:2018cvk}
T.~Bringmann and M.~Pospelov, \emph{{Novel direct detection constraints on light dark matter}}, \href{https://doi.org/10.1103/PhysRevLett.122.171801}{\emph{Phys. Rev. Lett.} {\bfseries 122} (2019) 171801} [\href{https://arxiv.org/abs/1810.10543}{{\ttfamily 1810.10543}}].

\bibitem{Agashe:2014yua}
K.~Agashe, Y.~Cui, L.~Necib and J.~Thaler, \emph{{(In)direct Detection of Boosted Dark Matter}}, \href{https://doi.org/10.1088/1475-7516/2014/10/062}{\emph{JCAP} {\bfseries 10} (2014) 062} [\href{https://arxiv.org/abs/1405.7370}{{\ttfamily 1405.7370}}].

\bibitem{Alvey:2020xsk}
J.~Alvey, N.~Sabti, V.~Tiki, D.~Blas, K.~Bondarenko, A.~Boyarsky et~al., \emph{{New constraints on the mass of fermionic dark matter from dwarf spheroidal galaxies}}, \href{https://doi.org/10.1093/mnras/staa3640}{\emph{Mon. Not. Roy. Astron. Soc.} {\bfseries 501} (2021) 1188} [\href{https://arxiv.org/abs/2010.03572}{{\ttfamily 2010.03572}}].

\bibitem{Boyarsky:2008ju}
A.~Boyarsky, O.~Ruchayskiy and D.~Iakubovskyi, \emph{{A Lower bound on the mass of Dark Matter particles}}, \href{https://doi.org/10.1088/1475-7516/2009/03/005}{\emph{JCAP} {\bfseries 03} (2009) 005} [\href{https://arxiv.org/abs/0808.3902}{{\ttfamily 0808.3902}}].

\bibitem{Boehm:2013jpa}
C.~Boehm, M.J.~Dolan and C.~McCabe, \emph{{A Lower Bound on the Mass of Cold Thermal Dark Matter from Planck}}, \href{https://doi.org/10.1088/1475-7516/2013/08/041}{\emph{JCAP} {\bfseries 08} (2013) 041} [\href{https://arxiv.org/abs/1303.6270}{{\ttfamily 1303.6270}}].

\bibitem{Sabti:2019mhn}
N.~Sabti, J.~Alvey, M.~Escudero, M.~Fairbairn and D.~Blas, \emph{{Refined Bounds on MeV-scale Thermal Dark Sectors from BBN and the CMB}}, \href{https://doi.org/10.1088/1475-7516/2020/01/004}{\emph{JCAP} {\bfseries 01} (2020) 004} [\href{https://arxiv.org/abs/1910.01649}{{\ttfamily 1910.01649}}].

\bibitem{Depta:2019lbe}
P.F.~Depta, M.~Hufnagel, K.~Schmidt-Hoberg and S.~Wild, \emph{{BBN constraints on the annihilation of MeV-scale dark matter}}, \href{https://doi.org/10.1088/1475-7516/2019/04/029}{\emph{JCAP} {\bfseries 04} (2019) 029} [\href{https://arxiv.org/abs/1901.06944}{{\ttfamily 1901.06944}}].

\bibitem{Soper:2014ska}
D.E.~Soper, M.~Spannowsky, C.J.~Wallace and T.M.P.~Tait, \emph{{Scattering of Dark Particles with Light Mediators}}, \href{https://doi.org/10.1103/PhysRevD.90.115005}{\emph{Phys. Rev. D} {\bfseries 90} (2014) 115005} [\href{https://arxiv.org/abs/1407.2623}{{\ttfamily 1407.2623}}].

\bibitem{Guo:2020oum}
G.~Guo, Y.-L.S.~Tsai, M.-R.~Wu and Q.~Yuan, \emph{{Elastic and Inelastic Scattering of Cosmic-Rays on Sub-GeV Dark Matter}}, \href{https://doi.org/10.1103/PhysRevD.102.103004}{\emph{Phys. Rev. D} {\bfseries 102} (2020) 103004} [\href{https://arxiv.org/abs/2008.12137}{{\ttfamily 2008.12137}}].

\bibitem{Chu:2020ysb}
X.~Chu, J.-L.~Kuo and J.~Pradler, \emph{{Dark sector-photon interactions in proton-beam experiments}}, \href{https://doi.org/10.1103/PhysRevD.101.075035}{\emph{Phys. Rev. D} {\bfseries 101} (2020) 075035} [\href{https://arxiv.org/abs/2001.06042}{{\ttfamily 2001.06042}}].

\bibitem{Alvey:2022pad}
J.~Alvey, T.~Bringmann and H.~Kolesova, \emph{{No room to hide: implications of cosmic-ray upscattering for GeV-scale dark matter}}, \href{https://doi.org/10.1007/JHEP01(2023)123}{\emph{JHEP} {\bfseries 01} (2023) 123} [\href{https://arxiv.org/abs/2209.03360}{{\ttfamily 2209.03360}}].

\bibitem{Donnachie:2002en}
S.~Donnachie, H.G.~Dosch, O.~Nachtmann and P.~Landshoff, \emph{{Pomeron physics and QCD}}, vol.~19, Cambridge University Press (12, 2004).

\bibitem{HESS}
\url{https://www.mpi-hd.mpg.de/HESS/}.

\bibitem{LHAASO}
\url{https://english.ihep.cas.cn/lhaaso}.

\bibitem{MAGIC}
\url{http://www.magic.iac.es}.

\bibitem{VERITAS}
\url{https://veritas.sao.arizona.edu}.

\bibitem{CTA}
\url{https://www.cta-observatory.org}.

\bibitem{CTAConsortium:2017dvg}
{\scshape CTA Consortium} collaboration, \emph{{Science with the Cherenkov Telescope Array}}, WSP (11, 2018), \href{https://doi.org/10.1142/10986}{10.1142/10986}, [\href{https://arxiv.org/abs/1709.07997}{{\ttfamily 1709.07997}}].

\bibitem{SWGO}
\url{https://www.swgo.org/SWGOWiki/doku.php}.

\bibitem{Cyburt:2002uw}
R.H.~Cyburt, B.D.~Fields, V.~Pavlidou and B.D.~Wandelt, \emph{{Constraining strong baryon dark matter interactions with primordial nucleosynthesis and cosmic rays}}, \href{https://doi.org/10.1103/PhysRevD.65.123503}{\emph{Phys. Rev. D} {\bfseries 65} (2002) 123503} [\href{https://arxiv.org/abs/astro-ph/0203240}{{\ttfamily astro-ph/0203240}}].

\bibitem{1996A&A...309..917A}
F.A.~{Aharonian} and A.M.~{Atoyan}, \emph{{On the emissivity of {\ensuremath{\pi}}\^0\^-decay gamma radiation in the vicinity of accelerators of galactic cosmic rays.}}, {\emph{Astron. Astrophys.} {\bfseries 309} (1996) 917}.

\bibitem{Stecker:1971ivh}
F.W.~Stecker, \emph{{Cosmic gamma rays}} (1971).

\bibitem{Gaisser:2016uoy}
T.K.~Gaisser, R.~Engel and E.~Resconi, \emph{{Cosmic Rays and Particle Physics}: {2nd Edition}}, Cambridge University Press (6, 2016).

\bibitem{2020MNRAS.495.4828G}
R.~{Guo}, C.~{Liu}, S.~{Mao} et~al., \emph{{Measuring the local dark matter density with LAMOST DR5 and Gaia DR2}}, \href{https://doi.org/10.1093/mnras/staa1483}{\emph{Mon. Not. Roy. Astron. Soc.} {\bfseries 495} (2020) 4828} [\href{https://arxiv.org/abs/2005.12018}{{\ttfamily 2005.12018}}].

\bibitem{HESS:2016pst}
{\scshape H.E.S.S.} collaboration, \emph{{Acceleration of petaelectronvolt protons in the Galactic Centre}}, \href{https://doi.org/10.1038/nature17147}{\emph{Nature} {\bfseries 531} (2016) 476} [\href{https://arxiv.org/abs/1603.07730}{{\ttfamily 1603.07730}}].

\bibitem{Gaggero:2017jts}
D.~Gaggero, D.~Grasso, A.~Marinelli, M.~Taoso and A.~Urbano, \emph{{Diffuse cosmic rays shining in the Galactic center: A novel interpretation of H.E.S.S. and Fermi-LAT $\gamma$-ray data}}, \href{https://doi.org/10.1103/PhysRevLett.119.031101}{\emph{Phys. Rev. Lett.} {\bfseries 119} (2017) 031101} [\href{https://arxiv.org/abs/1702.01124}{{\ttfamily 1702.01124}}].

\bibitem{Consolandi:2016fhd}
{\scshape AMS} collaboration, \emph{{Precision Measurement of the Proton Flux in Primary Cosmic Rays from 1 GV to 1.8 TV with the Alpha Magnetic Spectrometer on the International Space Station}},  in \emph{{25th European Cosmic Ray Symposium}}, 12, 2016 [\href{https://arxiv.org/abs/1612.08562}{{\ttfamily 1612.08562}}].

\bibitem{DAMPE:2023pjt}
{\scshape DAMPE} collaboration, \emph{{Measurement of the cosmic p+He energy spectrum from 46 GeV to 316 TeV with the DAMPE space mission}},  \href{https://arxiv.org/abs/2304.00137}{{\ttfamily 2304.00137}}.

\bibitem{Collins:1977jy}
P.D.B.~Collins, \emph{{An Introduction to Regge Theory and High-Energy Physics}}, Cambridge Monographs on Mathematical Physics, Cambridge Univ. Press, Cambridge, UK (5, 2009), \href{https://doi.org/10.1017/CBO9780511897603}{10.1017/CBO9780511897603}.

\bibitem{Menon:2013vka}
M.J.~Menon and P.V.R.G.~Silva, \emph{{A study on analytic parametrizations for proton\textendash{}proton cross-sections and asymptotia}}, \href{https://doi.org/10.1088/0954-3899/40/12/125001}{\emph{J. Phys. G} {\bfseries 40} (2013) 125001} [\href{https://arxiv.org/abs/1305.2947}{{\ttfamily 1305.2947}}].

\bibitem{Broilo:2020pjr}
M.~Broilo, V.P.~Gon\c{c}alves and P.V.R.G.~Silva, \emph{{Model of diffractive excitation in $pp$ collisions at high energies}}, \href{https://doi.org/10.1103/PhysRevD.101.074034}{\emph{Phys. Rev. D} {\bfseries 101} (2020) 074034} [\href{https://arxiv.org/abs/2003.04768}{{\ttfamily 2003.04768}}].

\bibitem{Fermi-LAT:2016zaq}
{\scshape Fermi-LAT} collaboration, \emph{{Development of the Model of Galactic Interstellar Emission for Standard Point-Source Analysis of Fermi Large Area Telescope Data}}, \href{https://doi.org/10.3847/0067-0049/223/2/26}{\emph{Astrophys. J. Suppl.} {\bfseries 223} (2016) 26} [\href{https://arxiv.org/abs/1602.07246}{{\ttfamily 1602.07246}}].

\bibitem{Navarro:1995iw}
J.F.~Navarro, C.S.~Frenk and S.D.M.~White, \emph{{The Structure of cold dark matter halos}}, \href{https://doi.org/10.1086/177173}{\emph{Astrophys. J.} {\bfseries 462} (1996) 563} [\href{https://arxiv.org/abs/astro-ph/9508025}{{\ttfamily astro-ph/9508025}}].

\bibitem{1965TrAlm...5...87E}
J.~{Einasto}, \emph{{On the Construction of a Composite Model for the Galaxy and on the Determination of the System of Galactic Parameters}}, {\emph{Trudy Astrofizicheskogo Instituta Alma-Ata} {\bfseries 5} (1965) 87}.

\bibitem{Burkert:1995yz}
A.~Burkert, \emph{{The Structure of dark matter halos in dwarf galaxies}}, \href{https://doi.org/10.1086/309560}{\emph{Astrophys. J. Lett.} {\bfseries 447} (1995) L25} [\href{https://arxiv.org/abs/astro-ph/9504041}{{\ttfamily astro-ph/9504041}}].

\bibitem{Montanari:2022buj}
A.~Montanari, E.~Moulin and N.L.~Rodd, \emph{{Toward the ultimate reach of current imaging atmospheric Cherenkov telescopes and their sensitivity to TeV dark matter}}, \href{https://doi.org/10.1103/PhysRevD.107.043028}{\emph{Phys. Rev. D} {\bfseries 107} (2023) 043028} [\href{https://arxiv.org/abs/2210.03140}{{\ttfamily 2210.03140}}].

\bibitem{Cirelli:2010xx}
M.~Cirelli, G.~Corcella, A.~Hektor, G.~Hutsi, M.~Kadastik, P.~Panci et~al., \emph{{PPPC 4 DM ID: A Poor Particle Physicist Cookbook for Dark Matter Indirect Detection}}, \href{https://doi.org/10.1088/1475-7516/2012/10/E01}{\emph{JCAP} {\bfseries 03} (2011) 051} [\href{https://arxiv.org/abs/1012.4515}{{\ttfamily 1012.4515}}].

\bibitem{Cautun:2019eaf}
M.~Cautun, A.~Benitez-Llambay, A.J.~Deason, C.S.~Frenk, A.~Fattahi, F.A.~G\'omez et~al., \emph{{The Milky Way total mass profile as inferred from Gaia DR2}}, \href{https://doi.org/10.1093/mnras/staa1017}{\emph{Mon. Not. Roy. Astron. Soc.} {\bfseries 494} (2020) 4291} [\href{https://arxiv.org/abs/1911.04557}{{\ttfamily 1911.04557}}].

\bibitem{Hopkins:2017ycn}
P.F.~Hopkins et~al., \emph{{FIRE-2 Simulations: Physics versus Numerics in Galaxy Formation}}, \href{https://doi.org/10.1093/mnras/sty1690}{\emph{Mon. Not. Roy. Astron. Soc.} {\bfseries 480} (2018) 800} [\href{https://arxiv.org/abs/1702.06148}{{\ttfamily 1702.06148}}].

\bibitem{McKeown:2021sob}
D.~McKeown, J.S.~Bullock, F.J.~Mercado, Z.~Hafen, M.~Boylan-Kolchin, A.~Wetzel et~al., \emph{{Amplified J-factors in the Galactic Centre for velocity-dependent dark matter annihilation in FIRE simulations}}, \href{https://doi.org/10.1093/mnras/stac966}{\emph{Mon. Not. Roy. Astron. Soc.} {\bfseries 513} (2022) 55} [\href{https://arxiv.org/abs/2111.03076}{{\ttfamily 2111.03076}}].

\bibitem{SWGOirf}
\url{https://github.com/harmscho/SGSOSensitivity}.

\bibitem{Aharonian:2009zk}
{\scshape H.E.S.S.} collaboration, \emph{{Spectrum and variability of the Galactic Center VHE gamma-ray source HESS J1745-290}}, \href{https://doi.org/10.1051/0004-6361/200811569}{\emph{Astron. Astrophys.} {\bfseries 503} (2009) 817} [\href{https://arxiv.org/abs/0906.1247}{{\ttfamily 0906.1247}}].

\bibitem{Aharonian:2008gw}
{\scshape H.E.S.S.} collaboration, \emph{{Exploring a SNR/Molecular Cloud Association Within HESS J1745-303}}, \href{https://doi.org/10.1051/0004-6361:20079230}{\emph{Astron. Astrophys.} {\bfseries 483} (2008) 509} [\href{https://arxiv.org/abs/0803.2844}{{\ttfamily 0803.2844}}].

\bibitem{HESS:2018pbp}
{\scshape HESS} collaboration, \emph{{The H.E.S.S. Galactic plane survey}}, \href{https://doi.org/10.1051/0004-6361/201732098}{\emph{Astron. Astrophys.} {\bfseries 612} (2018) A1} [\href{https://arxiv.org/abs/1804.02432}{{\ttfamily 1804.02432}}].

\bibitem{FermiLAT:2012aa}
{\scshape Fermi-LAT} collaboration, \emph{{Fermi-LAT Observations of the Diffuse Gamma-Ray Emission: Implications for Cosmic Rays and the Interstellar Medium}}, \href{https://doi.org/10.1088/0004-637X/750/1/3}{\emph{Astrophys. J.} {\bfseries 750} (2012) 3} [\href{https://arxiv.org/abs/1202.4039}{{\ttfamily 1202.4039}}].

\bibitem{Porter:2021tlr}
T.A.~Porter, G.~Johannesson and I.V.~Moskalenko, \emph{{The GALPROP Cosmic-ray Propagation and Nonthermal Emissions Framework: Release v57}}, \href{https://doi.org/10.3847/1538-4365/ac80f6}{\emph{Astrophys. J. Supp.} {\bfseries 262} (2022) 30} [\href{https://arxiv.org/abs/2112.12745}{{\ttfamily 2112.12745}}].

\bibitem{Montanari:2023sln}
A.~Montanari, O.~Macias and E.~Moulin, \emph{{TeV gamma-ray sensitivity to velocity-dependent dark matter models in the Galactic Center}}, \href{https://doi.org/10.1103/PhysRevD.108.083027}{\emph{Phys. Rev. D} {\bfseries 108} (2023) 083027} [\href{https://arxiv.org/abs/2309.09691}{{\ttfamily 2309.09691}}].

\bibitem{Adams:2021kwm}
{\scshape VERITAS} collaboration, \emph{{VERITAS Observations of the Galactic Center Region at Multi-TeV Gamma-Ray Energies}}, \href{https://doi.org/10.3847/1538-4357/abf926}{\emph{Astrophys. J.} {\bfseries 913} (2021) 115} [\href{https://arxiv.org/abs/2104.12735}{{\ttfamily 2104.12735}}].

\bibitem{MAGIC:2022acl}
{\scshape MAGIC} collaboration, \emph{{Search for Gamma-Ray Spectral Lines from Dark Matter Annihilation up to 100~TeV toward the Galactic Center with MAGIC}}, \href{https://doi.org/10.1103/PhysRevLett.130.061002}{\emph{Phys. Rev. Lett.} {\bfseries 130} (2023) 061002} [\href{https://arxiv.org/abs/2212.10527}{{\ttfamily 2212.10527}}].

\bibitem{Aharonian_2021}
{\scshape LHAASO} collaboration, \emph{Observation of the crab nebula with lhaaso-km2a - a performance study}, \href{https://doi.org/10.1088/1674-1137/abd01b}{\emph{Chinese Physics C} {\bfseries 45} (2021) 025002}.

\bibitem{PhysRevLett.129.261103}
{\scshape LHAASO} collaboration, \emph{Constraints on heavy decaying dark matter from 570 days of lhaaso observations}, \href{https://doi.org/10.1103/PhysRevLett.129.261103}{\emph{Phys. Rev. Lett.} {\bfseries 129} (2022) 261103}.

\bibitem{CTAOirfprod3}
C.T.A.~Observatory and C.T.A.~Consortium, \emph{{CTAO Instrument Response Functions - version prod3b-v2}},  Aug., 2021.
\newblock 10.5281/zenodo.5163273.

\bibitem{CTAprod5}
C.T.A.~Observatory and C.T.A.~Consortium, \emph{{CTAO Instrument Response Functions - prod5 version v0.1}},  Sept., 2021.
\newblock 10.5281/zenodo.5499840.

\bibitem{Conceicao:2023tfb}
{\scshape SWGO} collaboration, \emph{{The Southern Wide-field Gamma-ray Observatory}}, \href{https://doi.org/10.22323/1.444.0963}{\emph{PoS} {\bfseries ICRC2023} (2023) 963} [\href{https://arxiv.org/abs/2309.04577}{{\ttfamily 2309.04577}}].

\bibitem{Albert:2019afb}
A.~Albert et~al., \emph{{Science Case for a Wide Field-of-View Very-High-Energy Gamma-Ray Observatory in the Southern Hemisphere}},  \href{https://arxiv.org/abs/1902.08429}{{\ttfamily 1902.08429}}.

\bibitem{Viana:2019ucn}
A.~Viana, H.~Schoorlemmer, A.~Albert, V.~de~Souza, J.P.~Harding and J.~Hinton, \emph{{Searching for Dark Matter in the Galactic Halo with a Wide Field of View TeV Gamma-ray Observatory in the Southern Hemisphere}}, \href{https://doi.org/10.1088/1475-7516/2019/12/061}{\emph{JCAP} {\bfseries 12} (2019) 061} [\href{https://arxiv.org/abs/1906.03353}{{\ttfamily 1906.03353}}].

\bibitem{Abdallah:2016ygi}
{\scshape H.E.S.S.} collaboration, \emph{{Search for dark matter annihilations towards the inner Galactic halo from 10 years of observations with H.E.S.S}}, \href{https://doi.org/10.1103/PhysRevLett.117.111301}{\emph{Phys. Rev. Lett.} {\bfseries 117} (2016) 111301} [\href{https://arxiv.org/abs/1607.08142}{{\ttfamily 1607.08142}}].

\bibitem{Abdallah:2018qtu}
{\scshape H.E.S.S.} collaboration, \emph{{Search for $\gamma$-Ray Line Signals from Dark Matter Annihilations in the Inner Galactic Halo from 10 Years of Observations with H.E.S.S.}}, \href{https://doi.org/10.1103/PhysRevLett.120.201101}{\emph{Phys. Rev. Lett.} {\bfseries 120} (2018) 201101} [\href{https://arxiv.org/abs/1805.05741}{{\ttfamily 1805.05741}}].

\bibitem{Cowan:2011an}
G.~Cowan, K.~Cranmer, E.~Gross and O.~Vitells, \emph{{Asymptotic formulae for likelihood-based tests of new physics}}, \href{https://doi.org/10.1140/epjc/s10052-011-1554-0}{\emph{European Physical Journal C} {\bfseries 71} (2011) 1554} [\href{https://arxiv.org/abs/1007.1727}{{\ttfamily 1007.1727}}].

\bibitem{Gluscevic:2017ywp}
V.~Gluscevic and K.K.~Boddy, \emph{{Constraints on Scattering of keV\textendash{}TeV Dark Matter with Protons in the Early Universe}}, \href{https://doi.org/10.1103/PhysRevLett.121.081301}{\emph{Phys. Rev. Lett.} {\bfseries 121} (2018) 081301} [\href{https://arxiv.org/abs/1712.07133}{{\ttfamily 1712.07133}}].

\bibitem{Xu:2018efh}
W.L.~Xu, C.~Dvorkin and A.~Chael, \emph{{Probing sub-GeV Dark Matter-Baryon Scattering with Cosmological Observables}}, \href{https://doi.org/10.1103/PhysRevD.97.103530}{\emph{Phys. Rev. D} {\bfseries 97} (2018) 103530} [\href{https://arxiv.org/abs/1802.06788}{{\ttfamily 1802.06788}}].

\bibitem{Slatyer:2018aqg}
T.R.~Slatyer and C.-L.~Wu, \emph{{Early-Universe constraints on dark matter-baryon scattering and their implications for a global 21 cm signal}}, \href{https://doi.org/10.1103/PhysRevD.98.023013}{\emph{Phys. Rev. D} {\bfseries 98} (2018) 023013} [\href{https://arxiv.org/abs/1803.09734}{{\ttfamily 1803.09734}}].

\bibitem{Buen-Abad:2021mvc}
M.A.~Buen-Abad, R.~Essig, D.~McKeen and Y.-M.~Zhong, \emph{{Cosmological constraints on dark matter interactions with ordinary matter}}, \href{https://doi.org/10.1016/j.physrep.2022.02.006}{\emph{Phys. Rept.} {\bfseries 961} (2022) 1} [\href{https://arxiv.org/abs/2107.12377}{{\ttfamily 2107.12377}}].

\bibitem{Collar:2018ydf}
J.I.~Collar, \emph{{Search for a nonrelativistic component in the spectrum of cosmic rays at Earth}}, \href{https://doi.org/10.1103/PhysRevD.98.023005}{\emph{Phys. Rev. D} {\bfseries 98} (2018) 023005} [\href{https://arxiv.org/abs/1805.02646}{{\ttfamily 1805.02646}}].

\bibitem{Super-Kamiokande:2022ncz}
{\scshape Super-Kamiokande} collaboration, \emph{{Search for Cosmic-Ray Boosted Sub-GeV Dark Matter Using Recoil Protons at Super-Kamiokande}}, \href{https://doi.org/10.1103/PhysRevLett.130.031802}{\emph{Phys. Rev. Lett.} {\bfseries 130} (2023) 031802} [\href{https://arxiv.org/abs/2209.14968}{{\ttfamily 2209.14968}}].

\bibitem{Daci:2015hca}
N.~Daci, I.~De~Bruyn, S.~Lowette, M.H.G.~Tytgat and B.~Zaldivar, \emph{{Simplified SIMPs and the LHC}}, \href{https://doi.org/10.1007/JHEP11(2015)108}{\emph{JHEP} {\bfseries 11} (2015) 108} [\href{https://arxiv.org/abs/1503.05505}{{\ttfamily 1503.05505}}].

\bibitem{Murgia:2017lwo}
R.~Murgia, A.~Merle, M.~Viel, M.~Totzauer and A.~Schneider, \emph{{''Non-cold'' dark matter at small scales: a general approach}}, \href{https://doi.org/10.1088/1475-7516/2017/11/046}{\emph{JCAP} {\bfseries 11} (2017) 046} [\href{https://arxiv.org/abs/1704.07838}{{\ttfamily 1704.07838}}].

\bibitem{Nadler:2019zrb}
E.O.~Nadler, V.~Gluscevic, K.K.~Boddy and R.H.~Wechsler, \emph{{Constraints on Dark Matter Microphysics from the Milky Way Satellite Population}}, \href{https://doi.org/10.3847/2041-8213/ab1eb2}{\emph{Astrophys. J. Lett.} {\bfseries 878} (2019) 32} [\href{https://arxiv.org/abs/1904.10000}{{\ttfamily 1904.10000}}].

\end{thebibliography}\endgroup

\end{document}